\documentclass[fleqn,usenatbib]{mnras}

\usepackage{newtxtext,newtxmath}

\usepackage[T1]{fontenc}

\DeclareRobustCommand{\VAN}[3]{#2}
\let\VANthebibliography\thebibliography
\def\thebibliography{\DeclareRobustCommand{\VAN}[3]{##3}\VANthebibliography}

\usepackage{graphicx}	
\usepackage{amsmath}	

\newcommand{\hMpc}{h^{-1}{\rm Mpc}}
\newcommand{\hMsun}{h^{-1}{\rm M}_\odot}

\title[A two-halo model for the galaxy-halo connection based on MTNG]{The MillenniumTNG Project: An improved two-halo model for the galaxy-halo connection of red and blue galaxies}

\author[B. Hadzhiyska et al.]{%
Boryana Hadzhiyska,$^{1,2,3}$\thanks{E-mail: boryana.hadzhiyska@cfa.harvard.edu}
Daniel Eisenstein$^{1}$,
Lars Hernquist$^{1}$,
R\"udiger Pakmor$^{5}$,
Sownak Bose$^{4}$,
\newauthor%
Ana Maria Delgado$^{1}$,
Sergio Contreras$^{6}$,
Rahul Kannan$^{1}$,
Simon D. M. White$^{5}$,
Volker Springel$^{5}$,
\newauthor%
Carlos Frenk$^{4}$,
C\'esar Hern\'andez-Aguayo$^{5,7}$,
Fulvio Ferlito$^{5}$,
and Monica Barrera$^{5}$
\\%
\\%
$^{1}$Harvard-Smithsonian Center for Astrophysics, 60 Garden St, Cambridge, MA 02138, USA\\%
$^{2}$Miller Institute for Basic Research in Science, University of California, Berkeley, CA, 94720, USA\\%
$^{3}$Physics Division, Lawrence Berkeley National Laboratory, Berkeley, CA 94720\\
$^{4}$Institute for Computational Cosmology, Department of Physics, Durham University, South Road, Durham, DH1 3LE, UK\\%
$^{5}$Max-Planck-Institut f\"{u}r Astrophysik, Karl-Schwarzschild-Str. 1, 85748, Garching, Germany\\%
$^{6}$Donostia International Physics Center (DIPC), Donostia-San Sebastian, Spain\\%
$^{7}$Excellence Cluster ORIGINS, Boltzmannstrasse 2, 85748 Garching, Germany
}

\date{Accepted XXX. Received YYY; in original form ZZZ}

\pubyear{2021}

\begin{document}
\label{firstpage}
\pagerange{\pageref{firstpage}--\pageref{lastpage}}
\maketitle

\begin{abstract}
Approximate methods to populate dark matter halos with galaxies are of great utility for cosmological inferences with large galaxy surveys. However, the limitations of simple halo occupation models (HODs) preclude a full use of small-scale galaxy clustering data and call for more sophisticated models that are augmented by additional parameters.  In this work, we study two galaxy populations, luminous red galaxies (LRGs) and star-forming emission-line galaxies (ELGs), at two distinct epochs, $z = 1$ and $z=0$, in the large volume, high-resolution hydrodynamical simulation of the MillenniumTNG project. In a previous study we concentrated on the small-scale, one-halo regime down to very small scales, $r \sim 0.1 \ \hMpc$, while here we focus on modeling galaxy assembly bias in the two-halo regime, $r \gtrsim 1 \ \hMpc$. Interestingly, the ELG signal exhibits scale dependence out to relatively large scales ($r \sim 20 \ \hMpc$), implying that the linear bias approximation for this tracer is invalid on these scales, contrary to common assumptions.  The 10-15\% discrepancy  present in the standard halo model prescription is only reconciled when we augment our halo occupation model with a dependence on extrinsic halo properties (``shear'' being the best-performing one) rather than intrinsic ones (e.g., concentration, or peak mass). We argue that this fact constitutes evidence for two-halo galaxy conformity. Including tertiary assembly bias (i.e.~a third property beyond mass and ``shear'') is not an essential requirement for reconciling the galaxy assembly bias signal of LRGs, but the combination of external and internal halo properties is still beneficial for recovering the ELG clustering. We also find that centrals in low-mass haloes dominate the assembly bias signal of the ELG and LRG populations. Finally, we explore the predictions of our model for higher-order galaxy statistics such as nearest-neighbour counts. The latter supplies additional information about galaxy assembly bias and can be used to break degeneracies between halo model parameters.
\end{abstract}

\begin{keywords}
large-scale structure of Universe -- galaxies: haloes -- cosmology: theory
\end{keywords}



\section{Introduction}
\label{sec:intro}

Cosmologists are increasingly faced with the task of making accurate predictions for the galaxy distribution in ever larger volumes, primarily in an effort to gain access to the invaluable linear modes that encode a trove of cosmological information. Along with this increase in volume, the precision of small-scale galaxy clustering measurements has improved dramatically, and provides another, less well-trodden path to constraining cosmology and to understanding astrophysical processes. But in order to stress-test and reliably extract constraints on the $\Lambda$CDM paradigm from small scales, cosmologists need to develop accurate models for the small-scale galaxy distribution as well.

There is a consensus that the most accurate way to do so involves \textit{ab-initio} computations such as hydrodynamical simulations, which meticulously trace and evolve the various components of the Universe that contribute to galaxy formation according to the governing physical laws. Unfortunately, current hydrodynamic galaxy formation simulations are unable to reach the volumes needed for a full analysis of  the data of observational cosmological surveys. Accordingly, cosmologists need to resort to approximate methods to model the connection between galaxies and the underlying matter distribution, allowing in that way an interpretation of the wealth of available and upcoming observational data.

One of the standard methods for analyzing small-scale galaxy clustering data from cosmological surveys involves equipping dark-matter-only simulations with some `galaxy-painting' technique that allows a statistical comparison of theoretical predictions to what is seen in surveys. Empirical models such as the halo occupation distribution \citep[HOD,][]{2002ApJ...575..587B,2002PhR...372....1C,2004MNRAS.350.1153Y,2005ApJ...633..791Z} and the subhalo abundance matching \citep[SHAM,][]{2006ApJ...647..201C,2010ApJ...717..379B,2014ApJ...783..118R,2016MNRAS.459.3040G,2016MNRAS.460.3100C} techniques offer a simple and computationally inexpensive approach for modelling  galaxy clustering by characterizing the relation between galaxies and host (sub)haloes. In this way, they allow the construction of a large number of mock catalogues as needed for cosmological inference.

However, one of the main limitations of these empirical models is their handling of the effect of `galaxy assembly bias' \citep{Gao2005}. Galaxy assembly bias refers to a manifestation of a discrepancy between the actual distribution of galaxies and one inferred from dark matter halos using their present-day mass alone \citep[e.g.,][]{2007MNRAS.374.1303C}. Instead, additional halo properties, for example halo formation time, local environment, concentration, triaxiality, spin, or velocity dispersion need to be considered to describe the clustering correctly. Galaxy assembly bias originates from two effects: halo assembly bias and halo occupation variation. The former manifests itself as a difference in the halo clustering among halos of the same mass, but that differ by some secondary property \cite[e.g., formation time, concentration, spin, see also][]{2007MNRAS.377L...5G}, while the latter comes from the dependence of the halo occupancy (i.e., the number of galaxies per halo) on properties of the host halo other than its mass \citep[e.g.,][]{2018ApJ...853...84Z,2018MNRAS.480.3978A}.

The standard implementation of the popular HOD model does not consider halo properties apart from mass, and hence completely neglects galaxy assembly bias. Similarly, the baseline SHAM model does not take baryonic effects such as tidal stripping and disruption into consideration, which affect subhalos in $N$-body and hydro simulations differently and may thus distort our ability to link subhalo properties between the two. The most straightforward versions of both approaches also fail to implement a dependence on environmental properties, which have recently been identified with a growing body of evidence \citep[e.g.,][]{2019arXiv190302007R, 2019arXiv190200030M, 2021MNRAS.501.1603H, 2020MNRAS.493.5506H}.

Several attempts have been made to incorporate assembly bias into the HOD framework \citep[e.g.,][]{2015MNRAS.454.3030P, 2016MNRAS.460.2552H, 2018MNRAS.478.2019Y, 2019MNRAS.488.3541W, 2019ApJ...872..115V}, most of which have augmented the model with a dependence on halo concentration (or closely related quantities). However, internal halo properties have been shown to be insufficient in reproducing the full galaxy assembly bias signal \citep[e.g.,][]{2007MNRAS.374.1303C,1911.02610,2020MNRAS.492.2739X}. External halo properties on the other hand, e.g., related to the halo local environment, appear to be able to account for the majority of galaxy assembly bias in hydrodynamic simulations and observations, and can be implemented as extensions to the HOD prescription \citep[e.g.,][]{2016arXiv160102693M, 2021MNRAS.502.3242X, 2021MNRAS.501.1603H, 2022MNRAS.510.3301Y}.

In this paper, we address the question of how to best model the occupation distributions of haloes according to the large hydrodynamical simulation of the MillenniumTNG (MTNG) project, for different redshifts and for different galaxy samples. In particular, we propose a simple and intuitive model for determining the occupation numbers of haloes and the distribution of satellites in halos, designed for creating realistic mock catalogues that can be used in observational analyses. In a previous companion paper \citep{Hadzhiyska+2022a}, we already addressed our modelling of the one-halo term at very small scales, while here we focus on the two-halo term. In particular, we test the ability of our new method to reproduce statistics of the galaxy distribution on small scales, $1 \, \hMpc \lesssim r \lesssim 30 \, \hMpc$, by fitting the free parameters of the model to the MTNG halo occupancy, and by making a prediction for the large-scale galaxy distribution. We ensure that the particular form and the halo properties we have chosen for our model are not artificially introducing a boost of the large-scale clustering, but have a physical meaning based on the halo occupancy. Through these predicted galaxy catalogues, we explore which and how many assembly bias properties are required to successfully recover the clustering of MTNG galaxies, and we speculate what the physical reasons behind the corresponding preferences are. Additionally, we explore alternative statistics to the standard two-point correlation function with the goal to better differentiate between the proposed models. Doing so with MTNG offers an important advantage over previous simulations, as MTNG's larger volume is crucial for accurately estimating higher-order statistics.

The outline of the paper is summarized as follows. In Section~\ref{sec:meth}, we introduce the MillenniumTNG simulation suite and the methods we adopt for selecting galaxies and defining halo properties. In Section~\ref{sec:res}, we discuss the predictions of our model for key galaxy summary statistics such as the two-point correlation function, the redshift-space clustering, and the nearest neighbour counts, and compare them with the ``truth'' according to MTNG. In Section~\ref{sec:conc}, we summarize our findings and give our conclusions.

\section{Methods}
\label{sec:meth}

\subsection{MillenniumTNG simulations}
\label{sec:mtng}

The simulation suite of the MillenniumTNG project consists of several hydrodynamical and $N$-body simulations of varying resolutions and box sizes, including also some simulations with a massive neutrino component. A detailed description of the full simulation set is given in \citet{Aguayo2022} and our further introductory papers for the project \citep{Pakmor2022, Kannan2022, Barrera2022, Bose2022, Delgado2022, Ferlito2022, Hadzhiyska+2022a}.

In this study, we employ the largest available full-physics simulation box and its dark matter only counterpart, containing $2 \times 4320^3$ and $4320^3$ resolution elements, respectively, in a comoving volume of $(500\,h^{-1} {\rm Mpc})^3$. These simulations use the same cosmological model as IllustrisTNG \citep{2017MNRAS.465.3291W,2018MNRAS.473.4077P,2018MNRAS.475..648P,2018MNRAS.475..624N,2018MNRAS.477.1206N,2018MNRAS.480.5113M,2018MNRAS.475..676S, 2019MNRAS.490.3234N,2019MNRAS.490.3196P}, and their resolution is comparable but slightly lower than that of the largest IllustrisTNG box, TNG300-1, with $2.1 \times 10^7 \hMsun$ for the baryons and $1.1 \times 10^8 \hMsun$ for the dark matter. In analogy with the naming conventions of IllustrisTNG, we refer to the hydrodynamic simulation as MTNG740 due to its boxsize of $L=500\,h^{-1}{\rm Mpc} = 738.12 \,{\rm Mpc}$, while for the dark matter only run we use MTNG740-DM. We note that 
\citet{Pakmor2022} show that the galaxy properties predicted by MTNG740 are generally remarkably consistent with those of TNG300, in some properties even with TNG100. To first order, MTNG740 can thus be viewed as extending the IllustrisTNG model to a volume nearly 15 times bigger while otherwise being very similar.

The present work is a follow-up study to our companion paper \citet[][hereafter Paper~I]{Hadzhiyska+2022a}.  As in this paper, we refer to the ``virial'' mass and virial radius of a halo as the mass and radius that encloses a spherical region around the halo center with an overdensity value relative to the critical density that is derived from a generalization of the top-hat collapse model to low-density cosmologies \citep{1998ApJ...495...80B}.

\subsection{Galaxy populations}
\label{sec:gal_sel}

Similarly to Paper I \citep{Hadzhiyska+2022a}, we extract in this work LRGs and ELGs at redshifts $z = 0$ and $z=1$, with two number densities $n_{\rm gal} = [7.0 \times 10^{-4}, 2.0 \times 10^{-3}] \, (\hMpc]^{-3})$ corresponding to $N_{\rm gal} = [87 \ 800, 250 \ 000]$ galaxies. For full details on the selection, we refer the reader to Section~2.2 of \citet{Hadzhiyska+2022a}. Throughout the present text, we use the shortcuts `low'' and ``high'' for the two considered number densities $n_{\rm gal}$, respectively.  

We assume that the halo occupation distribution of the red (LRG-like) sample is well-approximated by the empirical formula given in \citet{Zheng:2004id}, according to which the mean halo occupation of centrals, $\langle N_{{\rm cen}} \rangle$, and satellites, $\langle N_{{\rm sat}} \rangle$, as a function of mass, $M$, is: \begin{equation}\label{eq:ncen}
    \langle N_{{\rm cen}} (M) \rangle = \frac{1}{2} \left[ 1 + {\rm erf} \left( \frac{\log M-\log M_{{\rm min}}}{\sigma_{{\log M}}} \right) \right], 
\end{equation}
\begin{equation}\label{eq:nsat}
    \langle N_{{\rm sat}} (M)\rangle = \left( \frac{M-M_{{\rm cut}}}{M_1} \right)^\alpha \, ,
\end{equation}
where $M_{\rm min}$ is the characteristic minimum mass of halos that host central galaxies, $\sigma_{\log M}$ is the width of this transition, $M_{{\rm cut}}$ is the characteristic cut-off scale for hosting satellites, $M_1$ is a normalization factor, and $\alpha$ is the power-law slope. For the blue (ELG-like) sample, we adopt the High Mass Quenched (HMQ) model proposed in \citet{2020MNRAS.497..581A}, which expresses the mean central occupation as:
\begin{align}
    \left< N_{\rm cen}\right>(M) &=  2 A \phi(M) \Phi(\gamma M)  + & \nonumber \\  
    \frac{1}{2Q} & \left[1+{\rm erf}\left(\frac{\log_{10}{M}-\log_{10}{M_c}}{0.01}\right) \right],  \label{eq:NHMQ}\\
\phi(x) &=\mathcal{N}(\log_{10}{ M_c},\sigma_M), \label{eq:NHMQ-phi}\\
\Phi(x) &= \int_{-\infty}^x \phi(t) \, \mathrm{d}t = \frac{1}{2} \left[ 1+{\rm erf} \left(\frac{x}{\sqrt{2}} \right) \right], \label{eq:NHMQ-Phi}\\
A &=\frac{p_{\rm max}  -1/Q }{\max(2\phi(x)\Phi(\gamma x))}.
\label{eq:NHMQ-A}
\end{align}
The occupation statistics of the satellites are assumed to obey the standard functional form of Eq.~(\ref{eq:nsat}).

\subsection{Halo properties}
\label{sec:halo_par}

In this section, we review the definitions of the various halo properties we employ to study galaxy assembly bias. We base our choice of which parameters to consider  on previous results that have identified  the most promising ones for explaining the galaxy assembly bias of IllustrisTNG \citep{1911.02610,2021MNRAS.508..698H,2021arXiv211102422D}.

\subsubsection{Adaptive environment}
\label{sec:env}

Previous works using hydrodynamical simulations suggest that halo models that condition on ``environment'' (defined in various ways) as a secondary parameter exhibit a substantial increase in the galaxy clustering \citep{2021MNRAS.501.1603H,2020MNRAS.493.5506H}. This suggests that in addition to the well-established halo clustering dependence on ``environment,'' there is an additional strong correlation between ``environment'' and halo occupancy, which manifests itself in the halo occupation distribution of a galaxy sample \citep[see][for studies of the halo occupancy dependence on ``environment'']{sownak, 1911.02610, 2021MNRAS.502.3242X, 2021MNRAS.502.3582Y}. While this finding signifies that haloes residing in denser regions contain more galaxies on average than haloes in underdense regions at fixed mass, the explanation for this observation is not yet clear and could range from numerical artefacts related to the halo finding algorithm to astrophysical phenomena such as extended splashback, quenching, merger efficiency and gas feedback in dense and underdense regions \citep[e.g.,][]{2007MNRAS.378..641A, 2017A&A...598A.103P, 2018MNRAS.476.5442P, 2018MNRAS.473.2486S}.

We adopt the following definition of adaptive halo environment (``environment,'' for short) to assess its effects on the large-scale galaxy distribution of MillenniumTNG:
\begin{enumerate}
\item Evaluate the dark matter overdensity field, $\delta (\mathbf{x})$, using triangular-shaped-cloud (TSC) interpolation on a $1024^3$ cubic lattice of all dark matter particles. In TSC, the fraction of a particle's mass assigned to a given cell of size $\Delta x$ along one dimension is determined by the shape function: $S(x) = (1-|x|/\Delta x)/\Delta x$. Each cell has size of $\Delta x = 500/1024 \ \hMpc \approx 0.5 \ \hMpc$ in our binning application.
\item Smooth the density field with a Gaussian kernel on scales $R_{\rm smooth} = 1.1$, $1.5$, $2.0$,  and $2.5\,  \hMpc$.
\item The smoothing scale used for a given halo is selected based on the halo virial radius: i.e.~we choose the closest smoothing scale to the halo radius rounding up. Thus, we make sure to always define the ``environment'' of a given halo on scales larger than its virial radius.
\item The halo property we call ``environment'' is then determined by the value of the smoothed density field interpolated at the halo centre-of-potential.
\end{enumerate}

\subsubsection{Adaptive shear}
\label{sec:shear}

Our procedure for obtaining the adaptive halo shear (``shear,'' for short) is similar to the one we adopted for ``environment'', the only difference being that the smoothed density field is further manipulated into the shear field \citep{2021arXiv211102422D}. To calculate the local ``shear'' around a halo, we first compute a dimensionless version of the tidal tensor as $T_{ij}\equiv \partial^2 \phi_R/\partial x_i \partial x_j$, where $\phi_R$ is the dimensionless potential field calculated using Poisson's equation: $\nabla^2 \phi_R = -\rho_R/\bar{\rho}$ (the subscript $R$ corresponds to the choice of smoothing scale). We then calculate the tidal shear $q^2_R$ as:
\begin{equation}
 q^2_R\equiv \frac{1}{2} \big[ (\lambda_2-\lambda_1)^2+(\lambda_3-\lambda_1)^2+(\lambda_3-\lambda_2)^2\big] \, ,
\end{equation}
where $\lambda_i$ are the eigenvalues of $T_{ij}$. Physically, ``shear'' and ``environment'' quantify different properties of the dark matter field. While the density is a reflection of the local distribution of matter at a given redshift, the ``shear'' measures the amount of anisotropic pulling due to gravity at a given point in space. Both ``shear'' and ``environment'' are thus ``extrinsic'' parameters, meaning that they refer to the halo surroundings rather than intrinsic properties.

\subsubsection{Concentration}
\label{sec:concentration}

The link between halo concentration and accretion history has been studied extensively in the literature \citep{1997ApJ...490..493N, Wechsler:2001cs, 2014MNRAS.441..378L, 2016MNRAS.460.1214L}. For example, it is well established that mass and concentration exhibit a negative correlation, such that massive haloes tend to exhibit richer substructure and a more spatially spread out distribution of their subhalos. Moreover, it has been shown that recent merger activity induces dramatic changes in halo concentrations, and that these responses linger over a period of several dynamical times, corresponding to many Gyr \citep[see, e.g.,][]{2020MNRAS.498.4450W}. Relevant to galaxy assembly bias studies is the fact that halo concentration has a bearing on both the halo occupation distribution and the halo clustering
\citep[e.g.,][]{2001MNRAS.321..559B,2014MNRAS.441..378L,2015ApJ...799..108D,2014MNRAS.441.3359D,2018MNRAS.474.5143M}.

In this study, we adopt the following proxy for the concentration of each MTNG halo:
\begin{equation}
c = V_{\rm max}/V_{\rm vir},
\end{equation}
where $V_{\rm max}$ is the maximum circular velocity of the halo at the redshift of interest, and $V_{\rm vir}$ is defined as $\sqrt{GM_{\rm vir}/R_{\rm vir}}$, where $M_{\rm vir}$ is the virial halo mass as defined by citet{1998ApJ...495...80B}.

\subsubsection{Velocity anisotropy}
\label{sec:vani}

Recent studies indicate that velocity anisotropy correlates with the galaxy clustering as predicted in IllustrisTNG \citep[e.g.,][]{2010ApJ...708..469F,2021MNRAS.501.1603H, 2021MNRAS.508..698H}. In this paper, we test this finding in the much larger volume of MillenniumTNG through a more robust galaxy population scheme. Following \citet{1987gady.book.....B}, we define the velocity anisotropy as
\begin{equation}
    \beta = 1-\frac{\sigma_{\rm tan}^2}{2 \sigma_{\rm rad}^2},
\end{equation}
where $\sigma_{\rm tan}$ and $\sigma_{\rm rad}$ are the tangential and radial velocity dispersions, respectively. We calculate these quantities over all dark matter particles in the FoF halo by projecting the velocity of each particle along and perpendicular to the radial direction (defined with respect to the position of the particle with the minimum gravitational potential) and then computing the standard deviation of each component  \citep{2019arXiv190302007R}. Thus  $\beta$ depends on the shape of the halo and captures information from the full phase-space structure of the parent halo. Haloes that have undergone recent accretion events tend to have particles which exhibit higher tangential velocities ($\sigma_{\rm tan}$) due to deflections caused by gravity shortly before accretion. This makes the velocity anisotropy parameter  well correlated with the recent merger history of a halo as well as its local ``environment'' \citep{2009MNRAS.394.1825F, 2010MNRAS.401.2245F,2017MNRAS.469..594B, 2021MNRAS.508..698H}.

\subsubsection{Peak virial mass}
\label{sec:peak}

We define the peak halo mass, $M_{\rm peak}$, as the total mass of the particles within the virial radius of a halo at the time this mass peaks over the halo's entire history. Recent studies suggest that populating subhaloes based on their early history properties (e.g.~mass or velocity at time of infall, or at their peak values) leads to a better agreement with the observed galaxy clustering \citep[e.g.][]{2016MNRAS.460.3100C}. A reasonable explanation for this finding is that when haloes orbit near a more massive neighbour, the outer layers of their dark matter distribution is often stripped, while their tighter cores remain intact. Thus, their stellar-to-halo mass ratio increases as a result of the interaction with the more massive object, making their total halo mass  a poor predictor of stellar mass. In such situations, one expects that the peak mass would be a better marker of the amount of luminous, stellar material in a halo.

\subsubsection{Splashback radius}
\label{sec:r_splash}

The splashback radius, $R_{\rm sp}$, has been proposed as a more physically motivated definition of the halo boundary \citep{2014ApJ...789....1D, 2014JCAP...11..019A, 2015ApJ...810...36M} than the conventionally used halo radius based on a particular overdensity contrast, which does not necessarily correspond with a physical feature in the density profile of a halo or the dynamics of its particles \citep[e.g.,][]{2017ApJ...843..140D, 2020ApJS..251...17D}. On the other hand, the splashback radius is the radius that particles reach on their first orbit after infall (apocentre), and as such is directly connected to the particle dynamics by separating infalling from orbiting matter \citep[][]{2018MNRAS.473.2486S}.

Here, we employ the python package {\small COLOSSUS} \citep{2018ApJS..239...35D} to determine the splashback radius
based on a fitting function. As input the method uses the halo radius $R_{\rm 200m}$ and the accretion rate, defined as:
\begin{equation}
    \Gamma_{\rm dyn} (t) = \frac{\Delta \log(M)}{\Delta \log(a)} = \frac{\log[M(t)] - \log[M(t-t_{\rm dyn})]}{\log[a(t)] - \log[a(t-t_{\rm dyn})]} ,
\end{equation}
where 
$M \equiv M_{\rm 200m}$ (mass within 200 times the mean density of the Universe) and the subscript ``${\rm dyn}$'' refers to the dynamical time, defined as:
\begin{equation}
    t_{\rm dyn}(z) \equiv t_{\rm cross}(z) = \frac{2 R_\Delta}{v_\Delta} = 2^{3/2} t_H(z) \left( \frac{\rho_\Delta(z)}{\rho_c(z)}\right)^{-1/2},
\end{equation}
where $t_H(z) \equiv 1/H(z)$ is the Hubble time. 

\begin{figure}
\centering  
\includegraphics[width=0.48\textwidth]{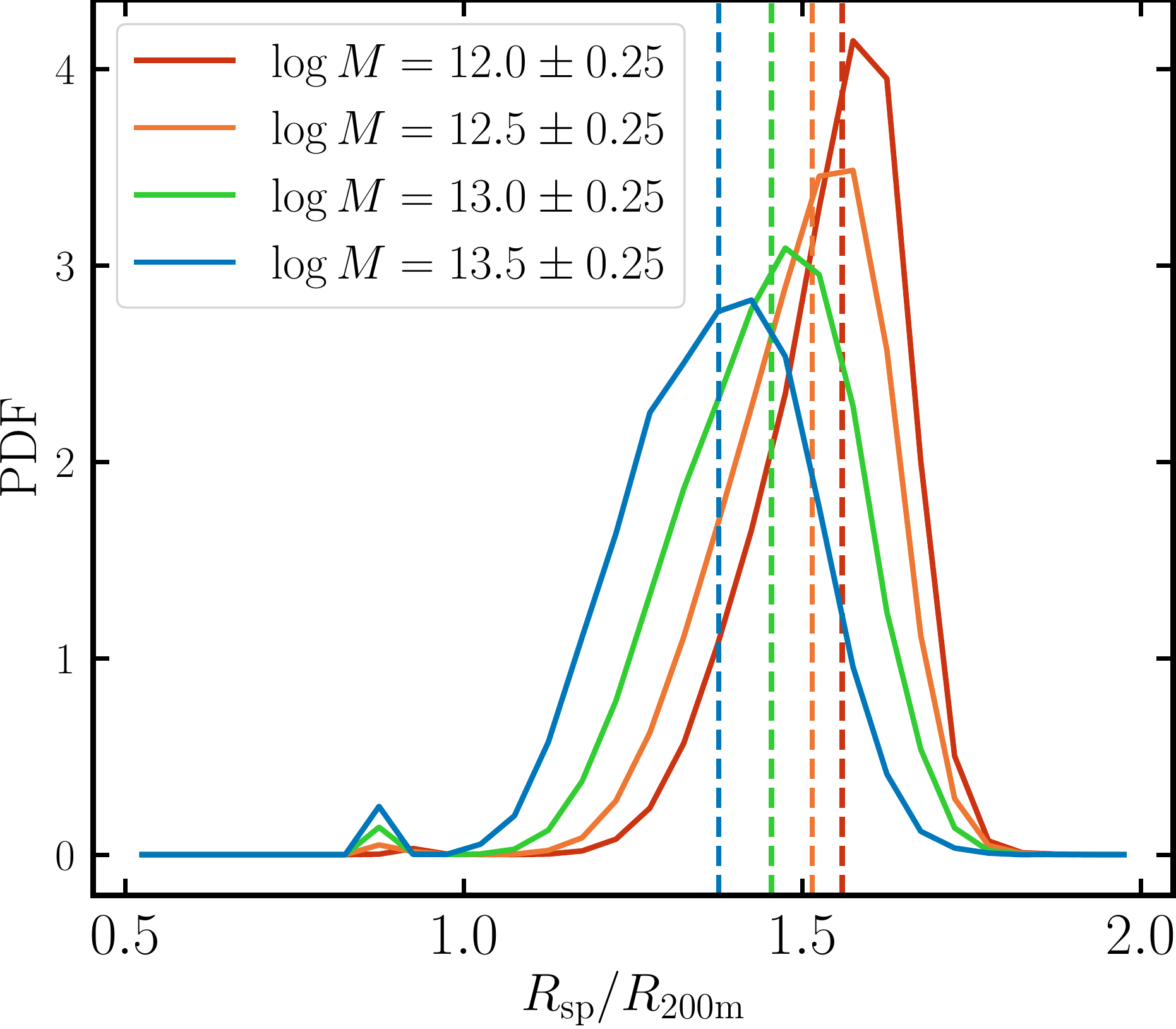}
\caption{Ratio of the empirically estimated splashback radius, $R_{\rm sp}$, based on the halo accretion history and the radius, $R_{\rm 200 m}$, of the haloes in the dark matter only MTNG box, at $z = 1$. On average, the splashback radius is about 50\% larger than $R_{\rm 200 m}$ for the four mass bins we consider. As we go to higher halo masses, the mean ratio $R_{\rm sp}/R_{\rm 200m}$ decreases, while the variance of the ratio increases.}
\label{fig:r_splash}
\end{figure}

In Fig.~\ref{fig:r_splash}, we show the ratio of the empirically estimated splashback radius, $R_{\rm sp}$, based on the halo accretion history, and the radius, $R_{\rm 200 m}$, at $z = 1$. Since the fitting function available in \textsc{colossus} loses accuracy for higher percentiles of the apocenter distribution, we adopt the largest credible splashback definition, corresponding to the 90th percentile of the particle distribution. The figure shows that the splashback radius is about 50\% larger than $R_{\rm 200 m}$ for the four mass bins we consider. This suggests that the radii and masses of haloes are underestimated when adopting $\Delta_{c,m}$-based definitions and that additional substructure is likely to reside outside the traditionally assumed halo boundary. This finding has important implications for the halo model, as a correction of the halo boundary based on the splashback radius would alter not only its mass, but also redefine the halo parentage of satellite galaxies. We explore this question in Section~\ref{sec:corr_splash}. Additionally, at higher halo masses, the mean ratio $R_{\rm sp}/R_{\rm 200m}$ decreases, while the variance of the ratio increases. Thus, the radii of smaller haloes are more likely to be appreciably underestimated, whereas in the case of larger haloes, the amount by which their radii are underestimated varies considerably. This is most likely a result of the overall more complicated dynamics of massive haloes, which are often located in denser regions and experience more mergers and other interactions with nearby haloes. (For a more detailed discussion of the splashback radius, see e.g., \citet{2020ApJS..251...17D,2020ApJ...903...87D}.)

\subsection{Nearest neighbour statistics}
\label{sec:kNN}

It is well known that the different galaxy samples targeted in spectroscopic surveys occupy different parts of the cosmic web \citep[][]{2018MNRAS.475.2530O, 2019MNRAS.483.4501A}. As a result, a single linear bias parameter cannot capture the full differences in the samples. The nearest neighbour framework, on the other hand, does recover the full statistics of the galaxy sample  \citep[e.g.,][]{2021MNRAS.500.5479B, 2022MNRAS.511.2765B}. As such, nearest neighbour measurements can be used to study assembly bias of different galaxy samples. Furthermore, adding nearest neighbour measurements to the data vectors used in the analysis of galaxy surveys can improve constraints on cosmological parameters even after marginalizing over all the galaxy-halo connection parameters.

To compute the nearest neighbour statistics of our sample, we follow the ``peaked kNN-CDF'' prescription from \citet{2021MNRAS.500.5479B}. We first generate a volume-filling sample of random points distributed over the total volume covered by the data, where the number of randoms is chosen to be larger than the number of data points to allow for better characterization of the tails of the distribution. For each random, we compute the distance to the $k$ nearest data point by constructing a tree using \texttt{scipy}'s  \texttt{cKDTree}, which returns a list of the distances to the $k$ nearest neighbours. Sorting these distances yields the Cumulative Distribution Function (CDF) in in the limit of a large number of randoms. Typically, we show the peaked kNN-CDF, defined as:
\begin{gather*}
\text{PCDF}(r) =      
\begin{cases}
  \text{CDF}(r) & \text{for CDF}(r) < 0.5 \\    
  1 - \text{CDF}(r) & \text{for CDF}(r) \geq 0.5,
\end{cases}
\end{gather*}
since it allows for better visual representation of the tails of the distribution. We note that for $k = 1$, $\text{VPF} = 1 - \text{CDF}(r)$ corresponds to the void probability function. For short, we refer to the peaked kNN-CDF as simply kNN. We compute the error bars on the kNN statistic using jackknife resampling by dividing the simulation volume into 125 equally sized subvolumes. We adopt a similar procedure when computing the error bars on the two-point correlation function in real space, and the monopole and quadrupole in redshift space, computed via the package \textsc{Corrfunc} \citep{2020MNRAS.491.3022S}.

\section{Results}
\label{sec:res}

In this section, we introduce an improved galaxy population model and compare its predictions with the true galaxy summary statistics obtained from the full-physics MTNG740 simulation. In particular, we fit our galaxy population model to the halo occupations of LRG- and ELG-hosting haloes in MTNG, and then test how well these fitted models can recover observable  large-scale structure statistics such as the two-point correlation function, redshift space distortions, and nearest neighbour counts. 

\subsection{Proposed halo occupation model}
\label{sec:halo_prop}

As demonstrated by many recent works \citep[e.g.,][]{sownak}, halo mass by itself is unable to recover the ``true'' occupancy of haloes in realistic galaxy models such as those provided by hydrodynamical simulations and semi-analytic models. Halo properties such as concentration, formation time and more recently, ``environment'' \citep[e.g.,][]{1911.02610, 2021MNRAS.502.3242X} have been shown to play a key role in determining the halo occupancy with higher accuracy. Despite their importance in recovering the ``true'' halo occupancy in realistic models, these parameters may not be good predictors for the true galaxy clustering. In this work, we explore which halo properties capable of recovering the halo occupancy in MTNG can also provide a galaxy sample that matches the clustering properties of galaxies in MTNG. To this end, we develop an intuitive model that extends the standard halo occupation model by adding two parameters for each assembly bias property.

Here we describe our galaxy-halo model for some galaxy tracer (e.g., ELGs, LRGs) in the case of two additional halo properties ($a$ and $b$), but the model is easily generalizable to three or more additional properties:
\begin{enumerate}
    \item First, we split the haloes with halo masses $\log(M) = 11-14$ into mass bins of size $\Delta \log(M) = 0.1$, measured in $\hMsun$. We leave the halo occupancies of the top 400 haloes, i.e., $\log(M) > 14$ at $z=1$ and $\log(M) > 14.3$ at $z = 0$, unmodified, as there are very few haloes in that mass regime so that  they might introduce a great deal of noise to our findings.
    \item For each halo in a given mass bin, we convert the values of the two halo properties of interest, $a$ and $b$, into two numbers, $f_a$ and $f_b$, by sorting, such that the halo with the lowest $a$ value gets $f_a = -0.5$, the highest gets $f_a = 0.5$, and the median gets $f_a = 0$ (and analogously for $b$). 
    \item We then model the halo occupancy by modifying the mean occupation number at fixed mass. This gives us a prediction for the number of centrals and satellites a given halo contains on average. The predicted halo occupancy of central galaxies for a given halo is given by:
    \begin{multline}
    \label{eq:ncen_ass}
        \hat N_{{\rm cen},i}(M, f_a, f_b) = \\ [1 + (a_{\rm cen} f_a + b_{\rm cen} f_b)(1-\langle N_{\rm cen} (M) \rangle)] \langle N_{\rm cen} (M) \rangle ,
    \end{multline}
    while in the case of satellites, it is:
    \begin{equation}
    \label{eq:nsat_ass}
        \hat N_{{\rm sat},i}(M, f_a, f_b) = [1 + (a_{\rm sat} f_a + b_{\rm sat} f_b)] \langle N_{\rm sat} (M) \rangle ,
    \end{equation}
    where $f_a$ and $f_b$ are the normalized rank-ordered halo properties, $a_{\rm cen}$ ($a_{\rm sat}$) and $b_{\rm cen}$ ($b_{\rm sat}$) are free parameters for the entire central (satellite) sample, and $\langle N_{\rm cen} (M) \rangle$ ($\langle N_{\rm sat} (M) \rangle$) are the mean number of centrals (satellites) in the mass bin of the halo under consideration. This value can either be taken directly from the true galaxy sample, or one can adopt an empirical formula such as Eqs.~\eqref{eq:ncen}, \eqref{eq:nsat}, and \eqref{eq:NHMQ}. 

    \textbf{Note 1:} The term $(1-\langle N_{\rm cen} (M) \rangle)$ is needed in order to ensure that as $\langle N_{\rm cen} (M) \rangle$ reaches its maximum value of one (typically for LRGs in the case of large halo masses), the modification to the occupancy becomes negligibly small. 

    \textbf{Note 2:} We require that $|a_{\rm cen, sat}| + |b_{\rm cen, sat}| \leq 2$ in order to ensure that the modified value of the mean occupancy is physical.
    We have tested different forms of Eq.~\eqref{eq:ncen_ass} that avoid this requirement. Examples of such forms are odd continuous functions that asymptote to a constant at some finite value of the argument such as hyperbolic tangent (${\rm tanh}(x)$), error function (${\rm erf}(x)$), Gudermannian function (${\rm gd}(x)$), or inverse tangent (${\rm arctan}(x)$). We find that ${\rm tanh}(x)$ gives almost indistinguishable results compared to Eq.~\eqref{eq:ncen_ass}, whereas the rest perform slightly worse. We select the linear form as easiest to interpret.

    \textbf{Note 3:} We choose the correction to be of the form $a_{\rm sat} f_a + b_{\rm sat} f_b$, after having carefully studied the three-dimensional surfaces of the simulation-predicted quantities $N_{\rm cen}(M, f_a, f_b)$ and $N_{\rm sat}(M, f_a, f_b)$ for each mass bin, and finding that they are well-approximated by hyper-planes with slopes $a_{\rm cen,sat}$ and $b_{\rm cen,sat}$ that are nearly constant over mass.
\end{enumerate}

Given the ``true'' occupancies of the haloes in MTNG, we can find the best-fitting values of the $a_{\rm cen,sat}$ and $b_{\rm cen,sat}$ parameters. The likelihood function to be maximised for the central occupations is provided by the binomial distribution (Bernoulli trials) as:
    \begin{equation}
        L(\hat N) = \prod_{i=1}^{N_{\rm h}} \hat N_i^{N_i} (1-\hat N_i)^{1-N_i} = \hat N_i^{N_{\rm tot}} (1-\hat N_i)^{N_{\rm h}-N_{\rm tot}} , 
    \end{equation}
where $\hat N_i \equiv \hat N_{{\rm cen},i}$, $N_i = \{ 0, 1\}$ is the ``true'' occupancy of centrals in MTNG, $N_{\rm tot}$ is the total number of centrals, and $N_{\rm h}$ is the total number of haloes. For the satellite occupations, we adopt the Poisson likelihood:
    \begin{equation}
        \log [L(\hat N)] = \sum_{i=1}^{N_{\rm h}} [N_i \log(\hat N_i) - \hat N_i], 
    \end{equation}
where $\hat N_i \equiv \hat N_{{\rm sat},i}$ and $N_i = \{ 0, 1\}$ is the ``true'' occupancy of satellites in MTNG. We find the best fit $a_{\rm cen,sat}$ and $b_{\rm cen,sat}$ parameters for a given galaxy sample via \texttt{scipy}'s implementation of the Nelder-Mead method. Substituting back the values of the free parameters into Eq.~\eqref{eq:ncen_ass} and Eq.~\eqref{eq:nsat_ass}, we obtain a prediction for the mean number of centrals and satellites expected in each halo. In the most vanilla scenario, we can turn the predicted mean occupancies into integers by drawing Bernoulli trials for the centrals, while for the satellites, we draw from a Poisson distribution. Alternatively, we can resort to using the correlated central-satellite occupation scheme described in Section 3.3, or the pseudo-Poisson distribution detailed in Section 3.2 of Paper I \citep{Hadzhiyska+2022a}.

\subsection{Galaxy assembly bias}
\label{sec:gab}

Galaxy assembly bias is the result of both halo assembly bias and occupancy variation, where the former is the dependence of the halo clustering on parameters other than halo mass, while the latter refers to the dependence of halo occupancy on halo properties other than halo mass. A standard way to measure galaxy assembly bias is by comparing the two-point correlation function of a galaxy sample with another sample, where for each halo, the galaxies populating it are randomly reassigned to another halo within the same mass bin \citep{2007MNRAS.374.1303C}. This ``random shuffling'' eliminates the dependence of halo occupancy on any secondary properties other than halo mass. If the ratio of the two-point correlation function of the shuffled to the original (unshuffled) sample differs from one, then the galaxy assembly bias signal is non-zero. This method of defining galaxy assembly bias is only meaningfully measurable in the two-halo regime; i.e. for $r \gtrsim 1 \ \hMpc$, as the scale $r \approx 1 \ \hMpc$ corresponds roughly to the transition regime between the one- and two-halo terms. We adopt the model for populating haloes with satellites described in Section 3.4 of Paper I \citep{Hadzhiyska+2022a}.

\begin{figure}
\centering  
\includegraphics[width=0.47\textwidth]{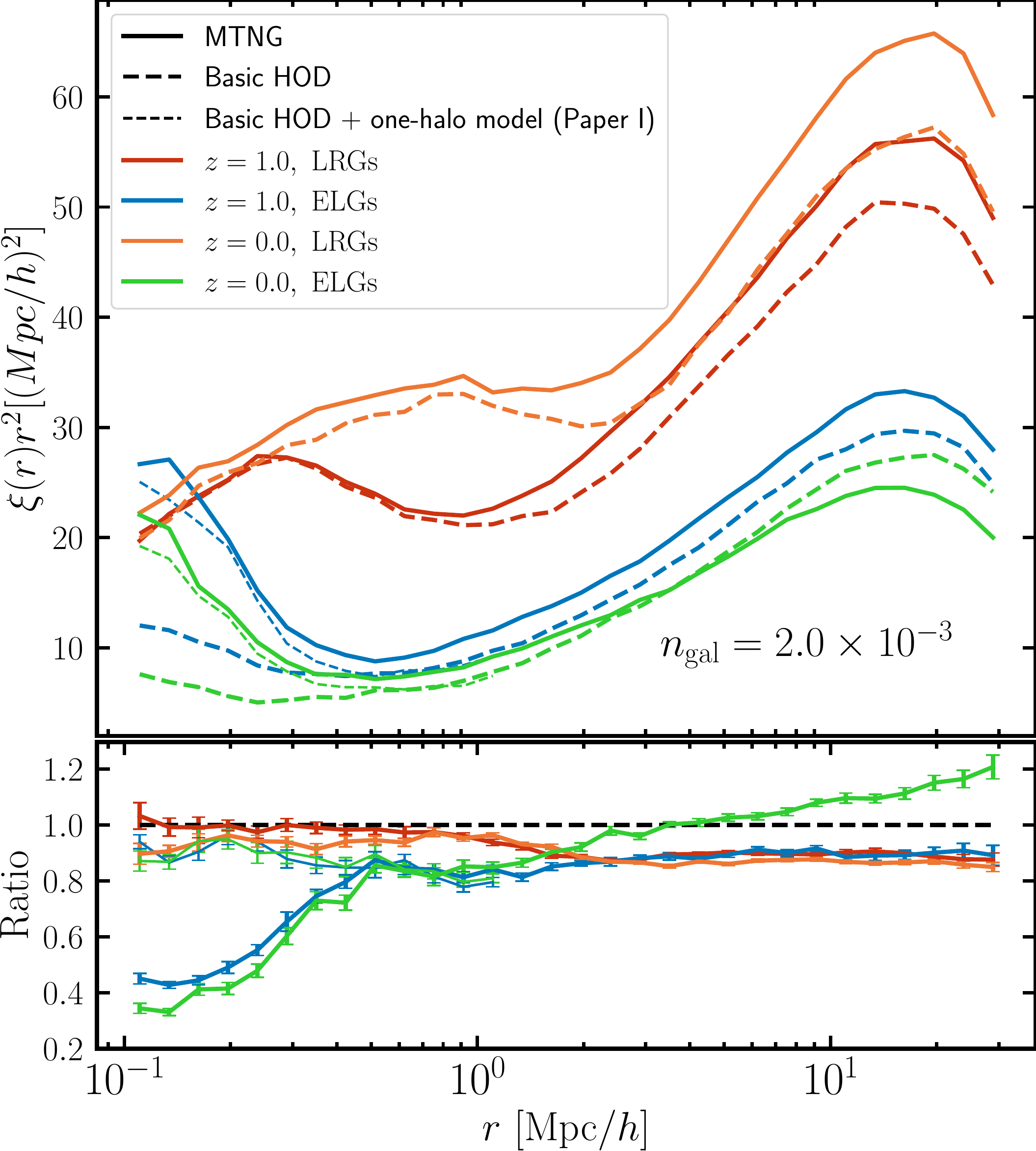}
\includegraphics[width=0.47\textwidth]{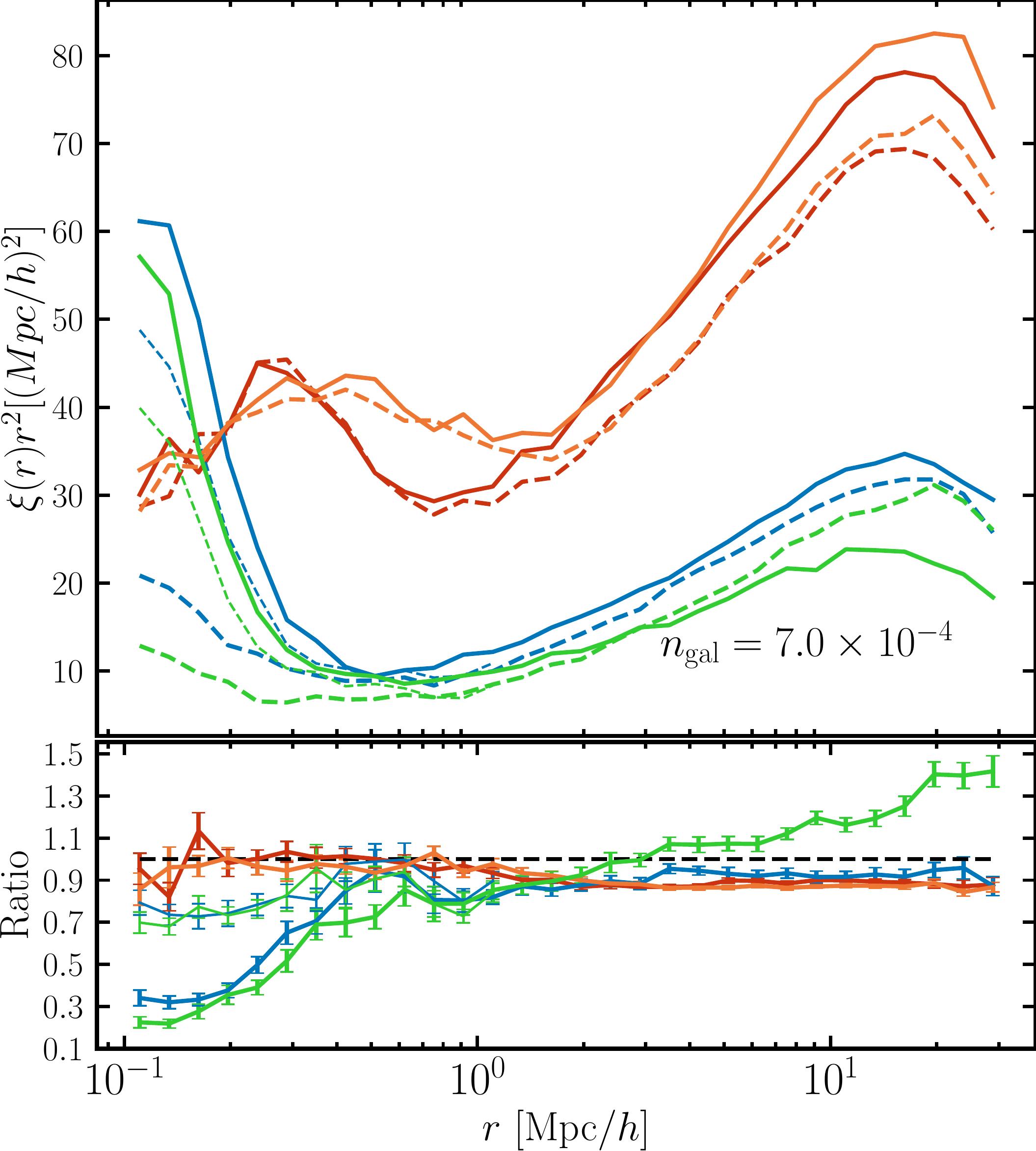}
\caption{Correlation function of the simple HOD (``predicted'') samples (dashed lines) and the ``true'' (solid lines) ``high''- (upper) and ``low''- (lower panel) density ELG and LRG samples extracted from MTNG740 at $z = 1$ and $z=0$. The lower segment shows the ratio of ``predicted'' to ``true'' clustering. The thin line shows the improved one-halo term ELG model prediction proposed in Paper~I \citep{Hadzhiyska+2022a}, incorporating pseudo-Poisson satellite statistics, and central-satellite and satellite-satellite galaxy conformity. For all samples but the $z = 0$ ELG one, the galaxy assembly bias signal is nearly constant, $\sim$5-10\%, indicating that at constant mass, galaxies prefer to live in more biased haloes. For the ELGs at $z = 0$, we conjecture that vigorous star formation happens only for central galaxies in underdense regions and newly merged satellites, hence the HOD sample is biased higher. Important for cosmological analyses is the scale-dependence of the ELG clustering, as it implies that linear bias is a poor approximation for that tracer.}
\label{fig:gab}
\end{figure}

In Fig.~\ref{fig:gab}, we show the correlation function of the predicted (i.e. adopting the ``shuffling'' technique) and the ``true'' (i.e. from MTNG) ``high''- and ``low''-density ELG and LRG samples at $z = 1$ and $z=0$. The shape of the correlation function is self-consistent for the two ELG samples, as well as for the two LRG samples at $z = 1$ and $z=0$, suggesting that as intended we are tracing similar populations at the two epochs. In particular, the large-scale clustering signal of the LRGs is stronger than that of the ELGs as they preferentially occupy more massive haloes, which tend to be more biased. Consequently, the one-halo to two-halo transition also occurs at larger scales for the LRGs. The ELG samples at both redshift epochs, on the other hand, have similar bias, as ELGs tend to be found in star-forming lower-mass haloes, which have not yet been quenched. As such, there is less redshift dependence in the selection of ELG-hosting haloes (and thus their bias). Interestingly, we notice that unlike the LRGs, on small scales, the ELGs appear to be much more clustered, which seems to confirm the hypothesis of cooperative galaxy formation, which manifests itself in the form of central-satellite and satellite-satellite conformity of the blue galaxies (see Paper I, \citet{Hadzhiyska+2022a}, for a more detailed discussion). This effect is even more pronounced in the ``low''-density regime, suggesting that the most star-bursty galaxies are the most likely ones to exhibit conformity, having had their star formation most recently triggered. This also agrees with the observed clustering of ELGs on small scales in the DESI survey\footnote{Personal correspondence with the DESI team.}. On the other hand, since the LRG hosts are the most massive haloes at any given epoch, it is no surprise that with time the hosts grow in mass and the sample becomes even more biased. 

Galaxy assembly bias manifests itself primarily in the two-halo term, so the relevant scales for studying its effects are $r \gtrsim 2 \ \hMpc$ for the LRGs and $r \gtrsim 0.8 \ \hMpc$ for the ELGs. The ``shuffled'' sample is equivalent to the basic HOD recipe as long as one matches the mean halo occupation of the full-physics simulation in the assumed shape of the HOD and the Poisson statistics of the satellite occupation holds reasonably well \citep[the effect of which are seen only in the one-halo term, see Paper I,][]{Hadzhiyska+2022a}. 

As seen in Fig.~\ref{fig:gab}, all samples except for the ELGs at $z = 0$ exhibit a galaxy assembly bias signal at a nearly constant level of $\sim$10\%, where the predicted sample is less clustered than the ``truth''. This suggests that at constant mass, the ``true'' galaxies prefer to live in more biased haloes. For the ELGs at $z = 0$, the picture is different. By $z = 0$, vigorously star forming regions have become much rarer, especially in dense regions, which have long been quenched due to the large number of mergers they have undergone. Thus, the only galaxies that are still star forming are central galaxies in underdense regions and smaller satellite subhaloes on their first infall. As a result, when we shuffle the occupations at fixed mass, we end up assigning galaxies to more biased haloes. The importance of detecting galaxy assembly bias is that it heavily implies that according to the MTNG galaxy formation model, one ought to include extensions to the halo occupation model (i.e.~expand beyond the mass-only assumption) when performing cosmological analysis of LRG and ELG clustering.

Noteworthy is the scale-dependence of ELG clustering that we notice in the lower segments of Fig.~\ref{fig:gab}, as it implies that linear bias is a poor approximation to the clustering on fairly large scales. This is particularly relevant for cosmological analyses probing the quasi-nonlinear regime because it challenges the assumption that those scales are unaffected by galaxy physics (pushing back the so-called ``non-local'' scale). We conjecture that this scale-dependence reflects the type of astrophysical conditions required to form ELGs. Namely, it is likely that ELGs prefer to form far from massive clusters, whose feedback processes and ejected gas quench the galaxies in a large radius nearby. This radius varies depending on the size and feedback strength of the cluster and hence, the signal becomes dependent on scale. We come back to this idea when we discuss environmental bias in Section~\ref{sec:gab_sec}. On average, this radius expands as a function of time, which explains why the effect is stronger at lower redshift. Note that the ELG sample at $z = 1$ also exhibits mild scale-dependence (also evident in the bottom left panel of Fig.~\ref{fig:gab_mass}), and we test our guess at $z = 0.5$, for which we also find strong scale-dependent signal. We note that the scale dependence of star-forming galaxies has previously been observed in \citet{2021MNRAS.506.3155J} albeit with a slightly different sample selection procedure and a different galaxy formation model. On yet larger scales, we expect the ratio to approach a constant as in the case of the LRG samples. These scales are not reachable by the volume of MTNG (the errorbars become too large for $r > 40 \ \hMpc$). Therefore, we plan to revisit the question of finding the scales at which the ELG ratio asymptotes to a constant in a future study, where we reliably emulate the MTNG model in a $(3\,{\rm Gpc})^3$ volume based on the MTNG-XXL simulation.

It is also remarkable how poorly the standard satellite population technique does in the case of the ELGs ($r \lesssim 1 \ \hMpc$), as seen in Fig.~\ref{fig:gab}. In this case, for the standard satellite population technique, we have assumed that the satellites are isotropically distributed in the halo, with the spherically averaged profile being taken directly from the full-physics run at fixed halo mass. Additionally, as traditionally done, the shuffling of centrals and satellites is done independently. As a result, we see a significant discrepancy at $r \sim 0.1 \ \hMpc$, where the ELGs are 3-5 times more clustered than the basic HOD prediction. Interestingly, this issue is almost completely overcome when we introduce our one-halo model from Paper I \citep{Hadzhiyska+2022a}, which takes into account the anisotropy of ELG satellites (i.e., their tendency to appear in the halo in pairs and triplets) as well as the central-satellite correlated occupation statistics. We note that the standard satellite population technique enjoys much more success with the LRGs, for which the small-scale clustering is fairly well matched under the assumption of spherical symmetry and central-satellite independence.

\subsubsection{Halo mass dependence}
\label{sec:gab_mass}

\begin{figure*}
\centering  
\includegraphics[width=0.95\textwidth]{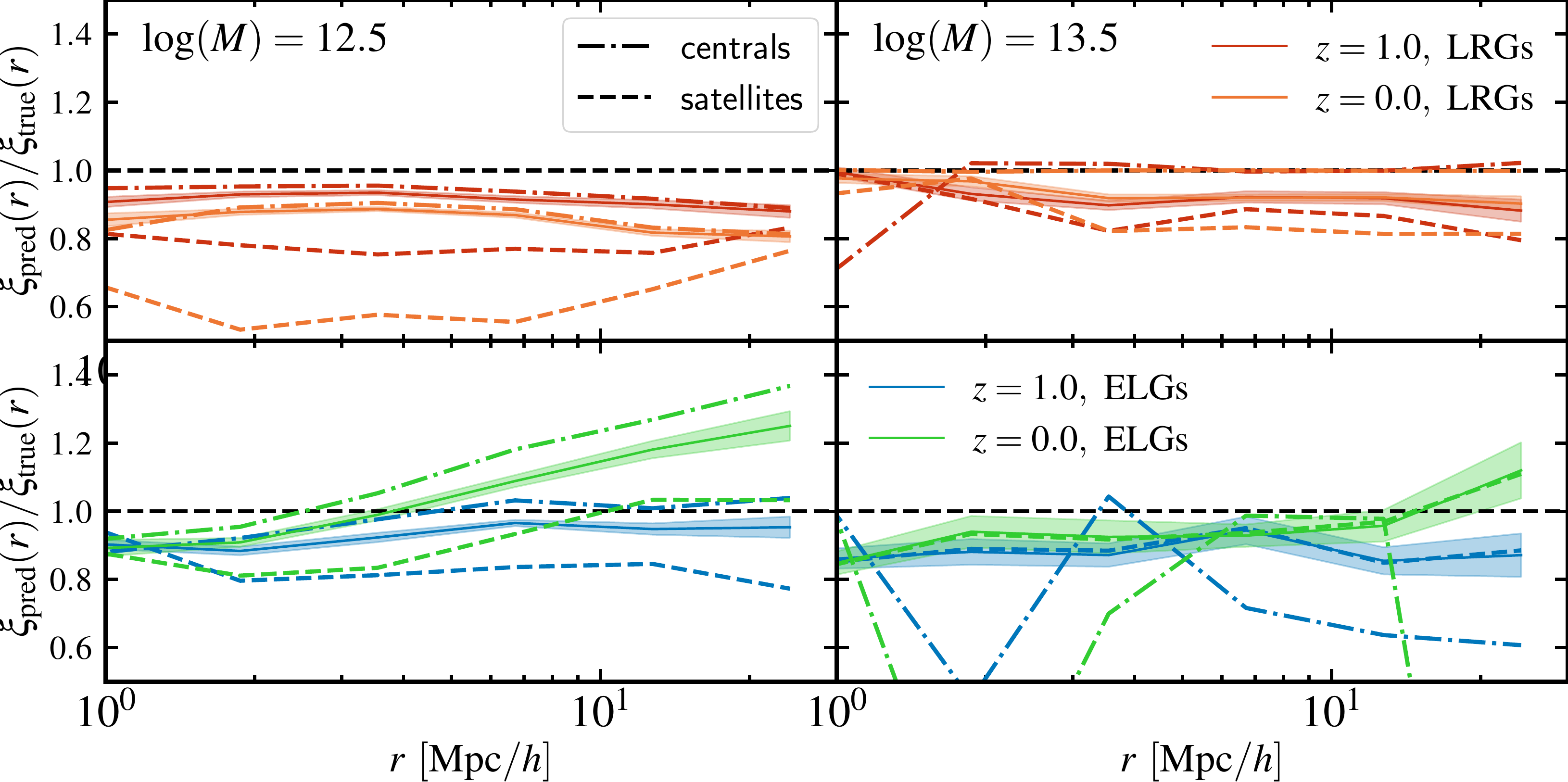}
\caption{Correlation function ratio of the predicted samples and the ``true'' ``high''-density ELG (lower panels) and LRG (upper panels) samples extracted from MTNG at $z = 1$ and $z=0$. Here, the predicted samples are obtained effectively via a mass-only HOD (i.e.~by shuffling the halo occupations in thin mass bins), and we aim to estimate the amount of galaxy assembly bias at different host halo masses (left panels: $M = 10^{12}-10^{13} \, \hMsun$; right panels: $M = 10^{13}-10^{14} \, \hMsun$). We split the contributions into the auto-correlation of all galaxies (thin line; shading denotes jackknife error bars), just centrals (dash-dotted line), and satellites (dashed line). We note that the centrals in the $\log[M/(  \hMsun )] = 12.5 \pm 0.5$ bin make up the largest fraction of the galaxies in each sample, and thus dominate the overall galaxy assembly bias signal. In the case of the LRGs, we see that the the central galaxies dictate the signal in the top left, whereas the satellites make the dominant contribution to the signal at higher masses, top right. In the ELG (lower) panel, the signal is again largely dictated by the centrals in the lower-mass bin, but the satellites also have a strong influence on it, as their fraction is larger (see text for fractions). For the higher-mass bin (lower right), we see that the satellites exhibit a moderate amount of galaxy assembly bias, though the precision is poorer due to their smaller number.}
\label{fig:gab_mass}
\end{figure*}

Next, we study the mass dependence of the galaxy assembly bias signal for the satellites and centrals, shown in Fig.~\ref{fig:gab_mass} as the correlation function ratio of the predicted samples and the ``true'' ``high''-density ELG and LRG samples at $z = 1$ and $z=0$. The predicted samples are obtained by shuffling the halo occupations in narrow mass bins as described in the previous section, and our goal is to estimate the amount of galaxy assembly bias at different masses. We split the galaxies into two groups, those hosted in haloes of mass $M = 10^{12}-10^{13}\,\hMsun$ and $M = 10^{13}-10^{14} \, \hMsun$. We study the auto-correlation ratios of all galaxies, just centrals, and the satellites. We note that the centrals in the lower-mass bin, i.e. $\log[M / (\hMsun)] = 12.5 \pm 0.5$, make up the largest fraction of the galaxies in each of the four samples. To quote numbers: for the ELGs, the satellite fraction is $\sim$36\% at both redshifts, and for the LRGs, it is $\sim$31\% at $z = 1$ and $\sim$39\% at $z = 0$. As a side note, the ``low''-density samples (not shown) have lower satellite fractions, but qualitatively the conclusions derived from the ``high'' density sample hold true. 

From Fig.~\ref{fig:gab_mass}, we see that the central LRG galaxies dictate the signal at lower-mass haloes, whereas the satellites have the dominant contribution at the higher-mass bin. At these higher masses, virtually all haloes have a LRG central, so there can be no galaxy assembly bias coming from the centrals. The overall galaxy assembly bias signal receives its largest contributions from low-mass centrals and thus, the biggest challenge is to find the astrophysical conditions (in the form of intrinsic and extrinsic halo properties) determining central occupancy at low masses. The ELG signal is also dominated by the centrals in the lower-mass bin, but the satellites also have a strong leverage on the overall signal, as their fractions are higher. In the higher-mass bin, we have very few central ELGs \citep[see Fig.~1 of Paper~I,][]{Hadzhiyska+2022a}, and the satellites take over the galaxy assembly bias signal. It is clear that the assembly bias properties of both centrals and satellites affect the accuracy of the mass-only halo model, albeit to a different extent and in different directions, and thus, these issues need to be addressed by a model that considers centrals and satellites separately.

\subsubsection{Secondary halo property dependence}
\label{sec:gab_sec}

In Section~\ref{sec:halo_prop}, we have proposed a model for determining the halo occupations, which builds on the traditional HOD approach by adding secondary (and tertiary) assembly bias parameters at the cost of two (or four) independent variables, $a_{\rm cen, sat}$ ($b_{\rm cen, sat}$), which modulate the occupation of the centrals and the satellites, respectively. In this section, we focus on the simpler version of our extended model; i.e.~including a dependence on a single halo property to account for secondary assembly bias. Our goal is to test which of the halo properties, if any, are capable of reconciling the difference in the two-halo regime, thus accounting for the galaxy assembly bias signal we observe in Fig.~\ref{fig:gab}. To this end, we fit the two parameters, $a_{\rm cen, sat}$, to the ``true'' \textit{halo occupations} extracted from MTNG for a galaxy sample of choice. We note that since our free parameters are fit to the halo occupations rather than the galaxy clustering, our procedure allows us to test consistently which of the six extended models (one for each of the halo properties in Section~\ref{sec:halo_par}) performs best; i.e.~is able to recover both the halo occupation statistics and make an accurate prediction of the large-scale clustering of galaxies. 

\begin{figure*}
\centering  
\includegraphics[width=0.97\textwidth]{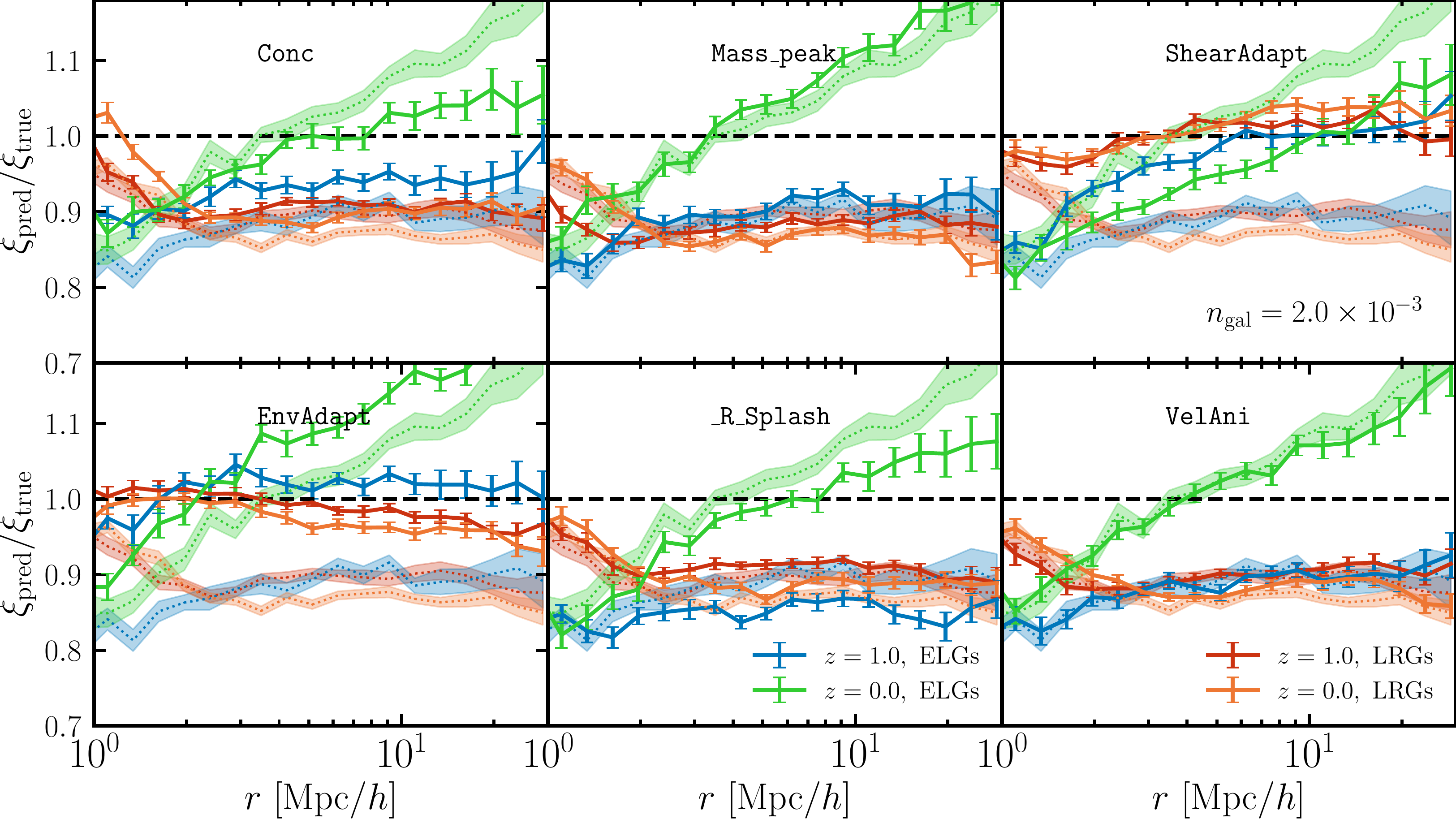}
\caption{Correlation function ratio of the predicted samples and the ``true'' ``high''-density ELG and LRG samples extracted from MTNG at $z = 1$ and $z=0$. Here, the predicted samples (solid lines with error bars) are obtained via our extended halo occupation model (see Section~\ref{sec:halo_prop}), where we have opted to include a dependence on a single additional halo property, thus exploring secondary assembly bias effects. The dotted shaded curves correspond to the mass-only HOD prediction (equivalent to the lower panel of Fig.~\ref{fig:gab}) and are repeated in each of the six panels for clarity. Each panel adopts a different halo property as a proxy of secondary assembly bias (see Section~\ref{sec:halo_par}). Relative to the mass-only HOD (dotted curves), the majority of the assembly-bias ``enhanced'' models shows some improvement. The halo property that appears to reconcile the difference in all four samples best is the ``shear''. This finding merits the attention of halo modeling prescriptions: the addition of two free parameters (tied to the halo ``shear''), $a_{\rm cen, sat}$ has the capacity to fully recover the real-space clustering of galaxies. Next best is ``environment'', which is not as successful with closing the ELG gap, but does an excellent job of reducing the large-scale discrepancy for the LRGs. We hypothesize that ``shear'' performs better than ``environment'' because it is defined via the gravitational potential rather than the smoothed density, and thus receives larger contributions from the halo surroundings compared with ``environment'', which is only sensitive to the immediate density at the halo boundary. The intrinsic halo properties (e.g., concentration, peak mass, splashback radius) have very little effect on the large-scale clustering of LRGs, but have more potential to improve the ELG auto-correlation. This suggests that ELG formation is more sensitive to internal halo factors.}
\label{fig:gab_ramp}
\end{figure*}

In Fig.~\ref{fig:gab_ramp}, we show our findings for the six extended models in the form of the correlation function ratio of the predicted samples and the ``true'' ``high''-density ELG and LRG samples at $z = 1$ and $z=0$. We also show the mass-only HOD prediction (Fig.~\ref{fig:gab}) to make the comparison easier. As in the previous figures we have shown, agreement of the ratio with unity signifies that the model provides a good match to the ``true'' galaxy clustering. As a reminder, the free parameters are not chosen to fit the clustering, but instead the halo occupation. We find that relative to the mass-only HOD, the majority of the enhanced models show some improvement. 

Intriguingly, the halo property that appears to best reconcile the deviation from one for all samples is the ``shear''. This is a promising finding that merits the attention of halo modeling prescriptions, as it implies that adding the minimum number of free parameters (tied to the halo ``shear''), $a_{\rm cen, sat}$ (see Section~\ref{sec:halo_prop} for a review of the model) to our halo occupation models has the capacity of fully recovering the real-space clustering of galaxies, according to the MTNG hydrodynamic simulation. The next best property is the ``environment'', which is not as successful with regards to the ELGs, but does an excellent job of reducing the large-scale discrepancy for the LRGs. The finding regarding LRGs echoes previous works \citep[e.g.,][]{1911.02610,2021MNRAS.502.3242X,2021MNRAS.502.3582Y}. We note that both of these properties are sensitive to the extrinsic properties of the halo, and as such are difficult to interpret from an astrophysical perspective. 

A plausible explanation for the environmental assembly bias of LRGs is cooperative galaxy formation; i.e.~the luminosity of a galaxy may be correlated on large scales with the luminosities of nearby galaxies as a result of proto-supercluster collapse \citep[see the discussion in Section~3.3 of Paper~I,][]{Hadzhiyska+2022a}. In the case of ELGs, we hypothesize that the reason ``shear'' performs better than ``environment'' is that the former is defined via the gravitational potential rather than the smoothed density, allowing it to receive larger-scale contributions from the halo surroundings, in contrast to ``environment,'' which is only sensitive to the immediate density at the halo boundary. For example, if there is a large cluster several megaparsecs away from a halo, ``environment'' would be completely agnostic, whereas ``shear'' could change in perceptible ways. This is in accordance with our conjecture from Fig.~\ref{fig:gab} that ELG formation may be anti-correlated with the presence of large clusters at very large distances ($\sim 10 \ \hMpc$). This phenomenon is referred to as two-halo galaxy conformity in the literature \citep{2015MNRAS.452.1958H, 2016MNRAS.461.2135H} and was first detected in observations and interpreted in \citet{2013MNRAS.430.1447K}.

It is noteworthy that the intrinsic halo properties (e.g., concentration, peak mass, splashback radius) have very little effect on the large-scale clustering of LRGs, but have a more noticeable effect on the ELG auto-correlation. This suggests that ELG formation is more sensitive to internal halo factors, whereas the LRGs depend on the large-scale placement of the halo with respect to other haloes and its interactions with them (e.g., mergers, disruptions, etc.). A few of the parameters appear to yield a worse agreement compared with the mass-only HOD for some samples (e.g. $R_{\rm splash}$ for $z = 1$, ELGs), hinting that a mass-independent central and satellite free parameter $a_{\rm cen, sat}$ (see Section~\ref{sec:halo_prop} for a review of the model) is unable to capture the evolution of the occupation statistics with redshift (and/or account for the dependence on the halo property under consideration). 

\subsubsection{Tertiary halo property dependence}
\label{sec:gab_thr}

Now that we have explored the performance of our extended halo occupation model in the event of augmenting it with a dependence on a single additional halo property, it is worth considering the effect of adding a second halo property. This exploration is important to the analysis of cosmological surveys, as it would help us to shed light on a long-standing question about the minimum number of ``nuisance'' parameters\footnote{The extended and traditional HOD parameters are considered nuisance parameters, since constraining them is rarely the end goal of any analysis.} that need to be included when analyzing observations in order to extract unbiased cosmological constraints. In addition, it is also of importance to galaxy formation, since it is possible that combinations of several intrinsic halo properties would capture the occupation statistics of haloes better than a single halo property. In the previous section, we found that both of the best-performing halo properties were extrinsic to the halo, ``shear'' and ``environment,'' defined on larger scales. Identifying intrinsic halo properties capable of reproducing the MTNG clustering would be valuable as these properties are much more readily tied to astrophysical phenomena and physical explanations.

\begin{figure*} \centering \includegraphics[width=0.97\textwidth]{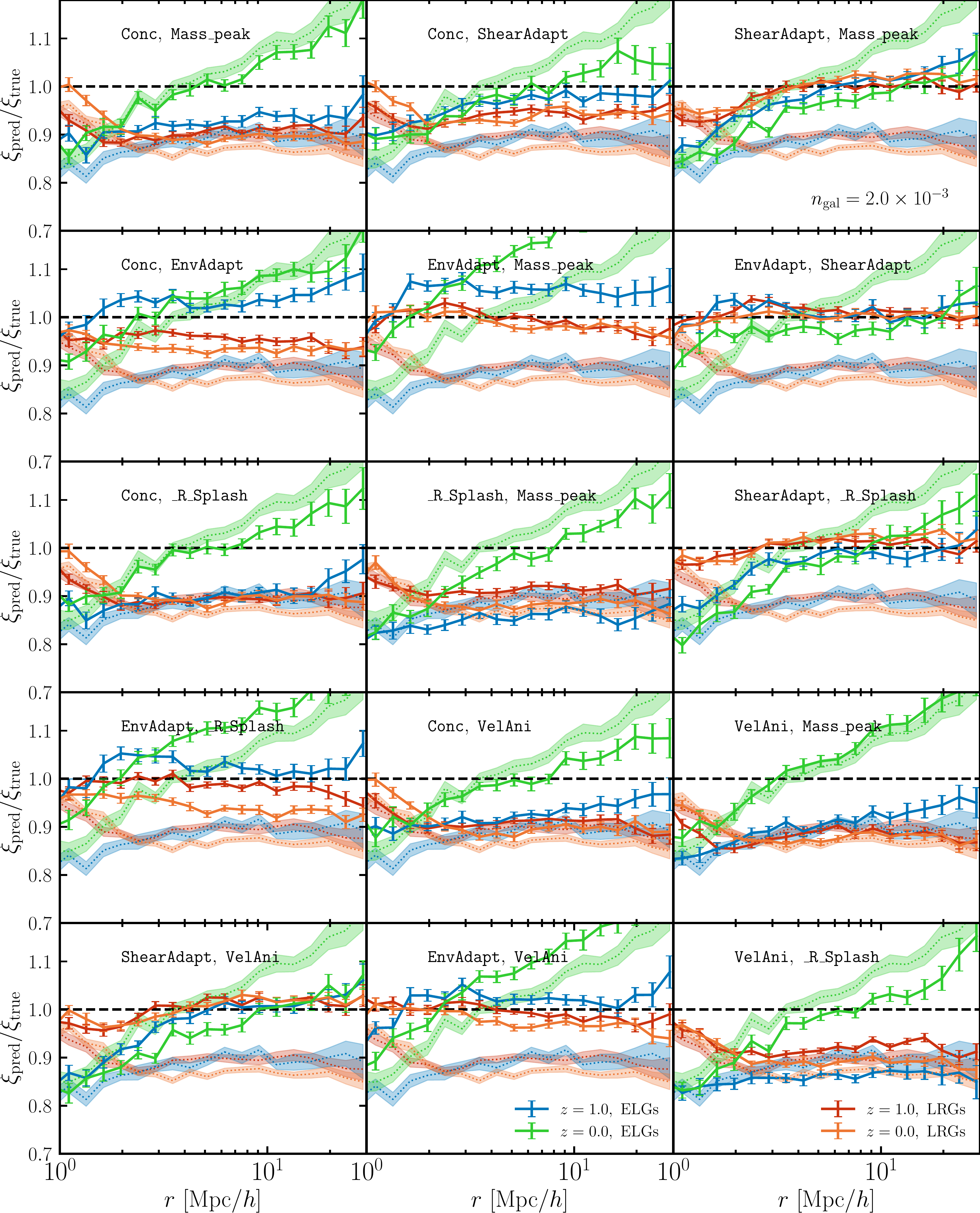} \caption{Correlation function ratio of the predicted samples and the ``true'' ``high''-density ELG and LRG samples at $z = 1$ and $z=0$. Here, the predicted samples (solid lines with error bars) are obtained via our extended halo occupation model (see Section~\ref{sec:halo_prop}), and we have opted to include a dependence on two additional halo properties (tertiary assembly bias).  The dotted shaded curves correspond to the mass-only HOD prediction and are repeated in each of the panels for clarity. Each panel adopts a different combination of halo properties for the secondary and tertiary assembly bias (see Section~\ref{sec:halo_par}).  The winning combination involves both ``environment'' and ``shear,'' demonstrating slightly better agreement with the full-physics catalog compared with ``shear''-only or ``environment''-only variants.}
\label{fig:gab_plane}
\end{figure*}

Fig.~\ref{fig:gab_plane} shows the correlation function ratio of the predicted samples and the ``true'' ``high''-density ELG and LRG samples extracted from MTNG at $z = 1$ and $z=0$. As in the previous section, the predicted samples are obtained via our extended halo occupation model for the two-halo term (see Section~\ref{sec:halo_prop}), where we have opted to include a dependence on two additional halo properties, thus testing the need for accounting for tertiary assembly bias. Complementing the findings in Fig.~\ref{fig:gab_ramp}, here, we explore 15 models enhanced with a different combination of halo properties for the secondary and tertiary assembly bias (see Section~\ref{sec:halo_prop} for a review of the model).

The best-performing models that use only intrinsic halo properties involve halo concentration and velocity anisotropy. However, the overall best models include either ``shear'' or ``environment''. We show that ``shear''-enhanced models show a lesser discrepancy in the case of the ELG samples, while ``environment''-enhanced models provide a better one-halo to two-halo transition in the case of the LRG samples. Thus, the optimal version of our extended HOD model adopts both ``environment'' and ``shear.'' We note that this combination demonstrates slightly better agreement with MTNG  compared with ``shear''-only or ``environment''-only (see Fig.~\ref{fig:gab_ramp}). The improvement is more notable for the ELG sample at $z = 0$. Combinations of extrinsic and intrinsic properties are also capable of reconciling the ELG clustering. For example, the velocity anisotropy plus shear model performs remarkably well for both ELGs and LRGs. If computational efficiency is a high priority to a pipeline analyzing cosmological data, we abstain from recommending to add a tertiary assembly bias enhancement to the HOD model in the case of LRGs, as those samples are particularly insensitive to the inclusion of a tertiary assembly bias property. However, there appears to be merit in including both intrinsic and extrinsic properties for recovering the clustering in the ELG case.

\subsubsection{Splashback radius correction}
\label{sec:corr_splash}

When adopting the splashback radius/mass definition in our analysis (Section~\ref{sec:r_splash}), we also reassign galaxies to the splashback-corrected FoF-identified haloes. To obtain the reassigned sample, we proceed as follows:
\begin{enumerate}
    \item We compute the splashback radii and masses (90-th percentile) of all haloes using the fitting functions in \textsc{COLOSSUS} given $R_{\rm 200m}$ and $\Gamma_{\rm dyn}$. 
    \item We rank order the haloes in terms of their splashback masses (starting with the highest), and for each, we find the galaxies residing within them. Once a galaxy has been assigned to a halo, it cannot be assigned to another one (thus, the preference is always to the higher-mass halo).
    \item The leftover galaxies ($\sim$5\%) are given to the haloes with highest enclosed density $M/R^3$ in their vicinity.
    \item We record the new halo parent of each galaxy, and the new occupancy number of each halo, which can then be used in the shuffling procedure.
\end{enumerate}

We next create two types of mock (or ``predicted'') galaxy catalogues using the procedure described above: one that adopts the splashback radius, and one that adopts the virial radius (defined using the generalized top-hat overdensity) as the halo boundary. This is done so that we can compare the two boundary definitions more directly. We note that this does not apply in the rest of the paper, where the halo occupations are determined based on the FoF boundaries of halos. The predicted samples are generated by randomly shuffling the satellite and central occupations of all haloes in a narrow mass bin of width $\Delta \log[M / (\hMsun)] = 0.1$. Once the number of satellites and centrals has been assigned (by shuffling the halo occupations at fixed mass bin), we compute the correlation function by giving a weight of $w = c_s+c_c$ to each halo, where $c_{s}$ and $c_c$ are the numbers of satellites and centrals in the halo, respectively. That way, we can isolate the effect on the two-halo term ($r \gtrsim 1 \ \hMpc$). 

\begin{figure}
\centering  
\includegraphics[width=0.48\textwidth]{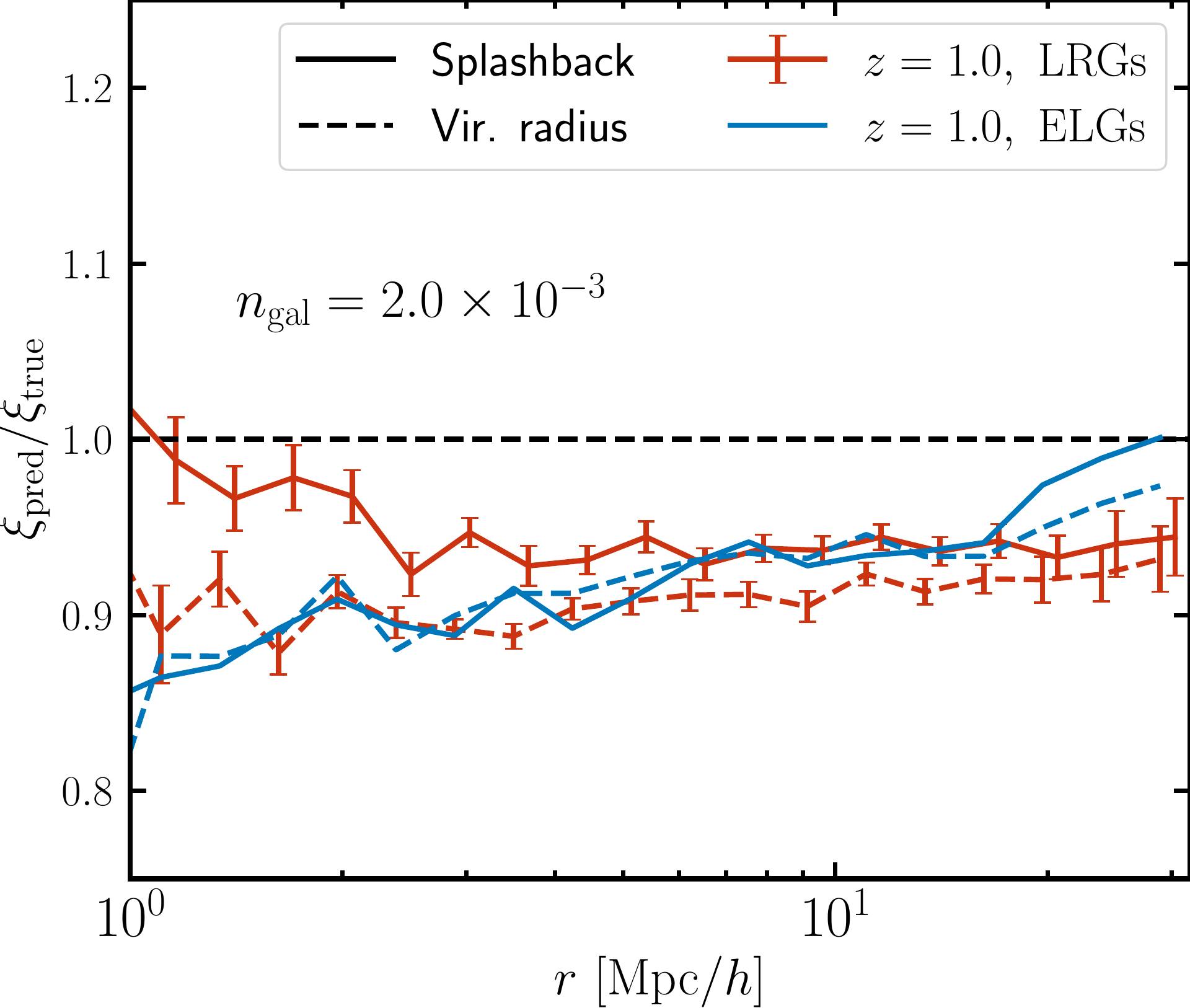}
\caption{Correlation function ratio between the predicted samples and the ``true'' LRG and ELG samples extracted from MTNG740 at $z = 1$. Results are shown for the ``high'' number density. The predicted samples are generated by randomly shuffling the satellite and central occupations of all haloes in a narrow mass bin of width $\Delta \log[M/(\hMsun)] = 0.1$, giving us effectively a mass-only HOD catalogue. The two-point correlation of the splashback-constructed ELG mock catalogue appears to be largely unchanged with respect to the virial radius-constructed mock, but in the case of the LRGs, the one-halo to two-halo transition regime ($1 \lesssim r \lesssim 3 \ \hMpc$) shows a smaller discrepancy with respect to the ``true'' LRG clustering. The two-halo term ($r \sim 1 \ \hMpc$) receives an improvement of 1-2\%. This shows that the splashback radius might be a more natural definition of the halo boundary.}
\label{fig:corr_splash}
\end{figure}

In Fig.~\ref{fig:corr_splash}, we show the ratio of the correlation function between the predicted and the ``true'' ``high''-density LRG and ELG samples extracted from MTNG at $z = 1$. The most significant contribution to the deviation from unity (5-10\%) that we see on large scales ($r \sim 10 \ \hMpc$) comes from the galaxy assembly bias effect (which is discussed in detail in Section~\ref{sec:gab}).  On smaller scales, the two-point correlation of the splashback-constructed ELG mock catalogue is largely unchanged with respect to the virial radius-constructed mock, but in the case of the LRGs, the splashback correction yields an improvement. Specifically, the one-halo to two-halo transition region ($1 \lesssim r \lesssim 3 \ \hMpc$) shows a smaller discrepancy with respect to the ``true'' LRG clustering, and the two are almost reconciled at the edge of the halo boundary ($r \sim 1 \ \hMpc$). The LRG clustering in the two-halo-regime ($r \sim 1 \ \hMpc$) receives a more modest, but consistent improvement of $\sim$2\%. The fact that the splashback mass results in a smaller discrepancy suggests that it is a more natural (and physical) proxy of halo mass. We conjecture that this is largely due to the boost in mass of small haloes near large ones, which makes it more likely for ``backsplash'' haloes to be assigned a galaxy, adding to the clustering near the halo boundary scale ($r \sim 1 \ \hMpc$).

A small additional effect that increases the overall bias comes from the reassignment of satellites, since our procedure preferentially assigns satellite occupations to the largest haloes, which have the highest bias. In other words, when we use the splashback definition vs. the virial radius one, we obtain a larger contribution to the signal from more biased haloes. Thus, the reassignment procedure modifies the halo occupation distribution by increasing the average number of satellites of high-mass haloes and decreasing that number in small haloes. In reality, the splashback radii of nearby haloes often overlap, and our choice of assigning the satellites in the overlapping regions to the largest halo is certainly an oversimplification. The more accurate approach would be to trace the trajectories of the satellite hosting subhaloes through time before deciding on their halo allegiance \citep[as done in e.g.][for the dark-matter particles]{sparta}. However, this is computationally extremely expensive, and would only make a small difference to the clustering, as the fraction of incorrectly assigned satellites is quite small. One simple way to test the magnitude of this effect is to try a hybrid version where we use the splashback mass as proxy when performing the shuffling, and the virial-radius assignment when ascribing the occupations. We find that this change makes negligible difference to the results at $z = 1$. We note that at lower redshifts, this effect matters more as the splashback radii overlap more due to the collapse of structure.

\subsection{Redshift-space distortions}
\label{sec:gab_rsd}

In the real Universe, we do not have the luxury of measuring the physical three-dimensional spatial coordinates of a galaxy. Instead, we can determine the two-dimensional galaxy position on the sky and the galaxy's redshift, which entangles its physical position with its velocity along the line of sight. As a result, any statistics that incorporates the along-the-line-of-sight direction inevitably gives us information about the peculiar velocities of galaxies. These statistics, if modelled well, are invaluable for cosmological studies, as they allow us to constrain the growth of structure via the infall velocities of galaxies. However, making such an inference comes with its own challenge, namely, the requirement to model galaxy peculiar velocities accurately.

\begin{figure*}
\centering  
\includegraphics[width=0.97\textwidth]{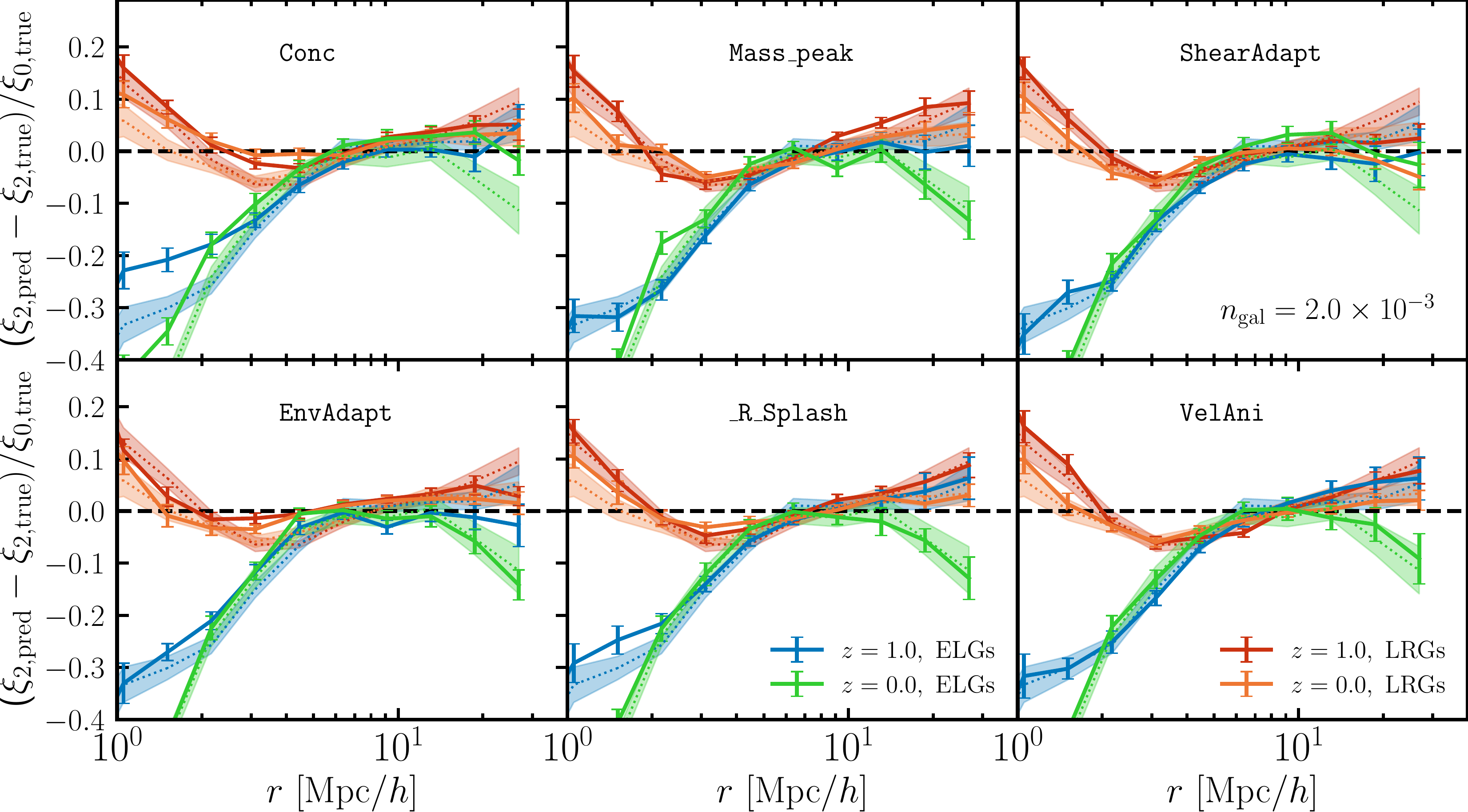}
\caption{Quadrupole moment difference between the predicted and ``true'' ``high''-density ELG and LRG samples at $z = 1$ and $z=0$. The predicted samples are obtained via our extended halo occupation model (see Section~\ref{sec:halo_prop}), and each panel adopts a different halo property as a secondary assembly bias proxy. Similarly to Fig.~\ref{fig:gab_ramp}, we find that the performance of the model enhanced with ``shear'' is the most optimal in the two-halo regime ($r \gtrsim 10 \ \hMpc$). As was the case with the real-space correlation function, ``environment'' performs well for all samples but the ELGs. None of the models enhanced with an intrinsic halo property (concentration, peak mass, velocity anisotropy, or splashback) can reconcile the large-scale ($r \gtrsim 10 \ \hMpc$) disagreement in the LRG curves, but some of them lead to an improvement in the ELG case (e.g., concentration and velocity anisotropy).}
\label{fig:xil2_ramp}
\end{figure*}

The statistics we analyze in this section are the monopole and quadrupole moments, $\xi_{\ell=0,2}(r)$, of the two-point function, which are the lowest order multipoles that capture the anisotropy due to redshift space distortions.  In Fig.~\ref{fig:xil2_ramp}, we compute the quadrupole moment for our proposed halo occupation model augmented with secondary assembly bias and show the difference $(\xi_{2, \rm pred}-\xi_{2, \rm true})/\xi_{0, \rm true}$. The convergence towards zero observed at $r \sim 10 \ \hMpc$ for all curves is due to the fact that $\xi_{\ell=2}(r)$ crosses zero around that scale, and thus the ratio to $\xi_{\ell=0}$ becomes very small. Similarly to Fig.~\ref{fig:gab_ramp}, we find that the performance of the model augmented with the halo property we call ``shear'' performs best in recovering the two-halo clustering of all MTNG galaxy samples (i.e., $r \gtrsim 10 \ \hMpc$). Similarly to the case of the real-space correlation function, the property ``environment'' is second-best in that regime, only struggling to recover the clustering of the ELGs.

None of the models enhanced with an intrinsic halo property (concentration, peak mass, velocity anisotropy, splashback) can reconcile the large-scale disagreement in the LRG curves, but some of them lead to an improvement in the ELG case (e.g., concentration and velocity anisotropy). In terms of the transitioning between one-halo and two-halo terms, it seems that the least discrepant predictions for all samples come from the ``concentration''-enhanced model. This is followed by halo ``shear'' and ``peak mass''. A general observation is that all models struggle with recovering the clustering on smaller scales ($r \sim 1 \ \hMpc$), and this is particularly true for the ELG curves. However, the intrinsic properties splashback radius, concentration, and peak mass, appear to help with the ELG samples, whereas the extrinsic ones appear to have almost no bearing on the one-halo term, which makes intuitive sense. A full analysis incorporating the one-halo model from Paper I \citep{Hadzhiyska+2022a} for the ELGs might significantly help with the one-halo to two-halo transition, but we leave this for future work, as it requires the development of a likelihood pipeline.

\subsection{Higher-order statistics: nearest neighbours}
\label{sec:gab_kNN}

In Section~\ref{sec:kNN}, we introduced an alternative summary statistics, which uses nearest-neighbour counts and may allow us to unlock valuable cosmological information encoded in the small-scale distribution of galaxies. Before this nearest-neighbour statistics can be reliably applied to real data, it needs to be robustly tested in simulations, and its sensitivity to both observational and astrophysical effects such as assembly bias needs to be understood well. In this work, we are interested in the latter question, which to our knowledge, has not yet been tackled through analyses involving hydrodynamical simulations.

\begin{figure*}
\centering  
\includegraphics[width=0.97\textwidth]{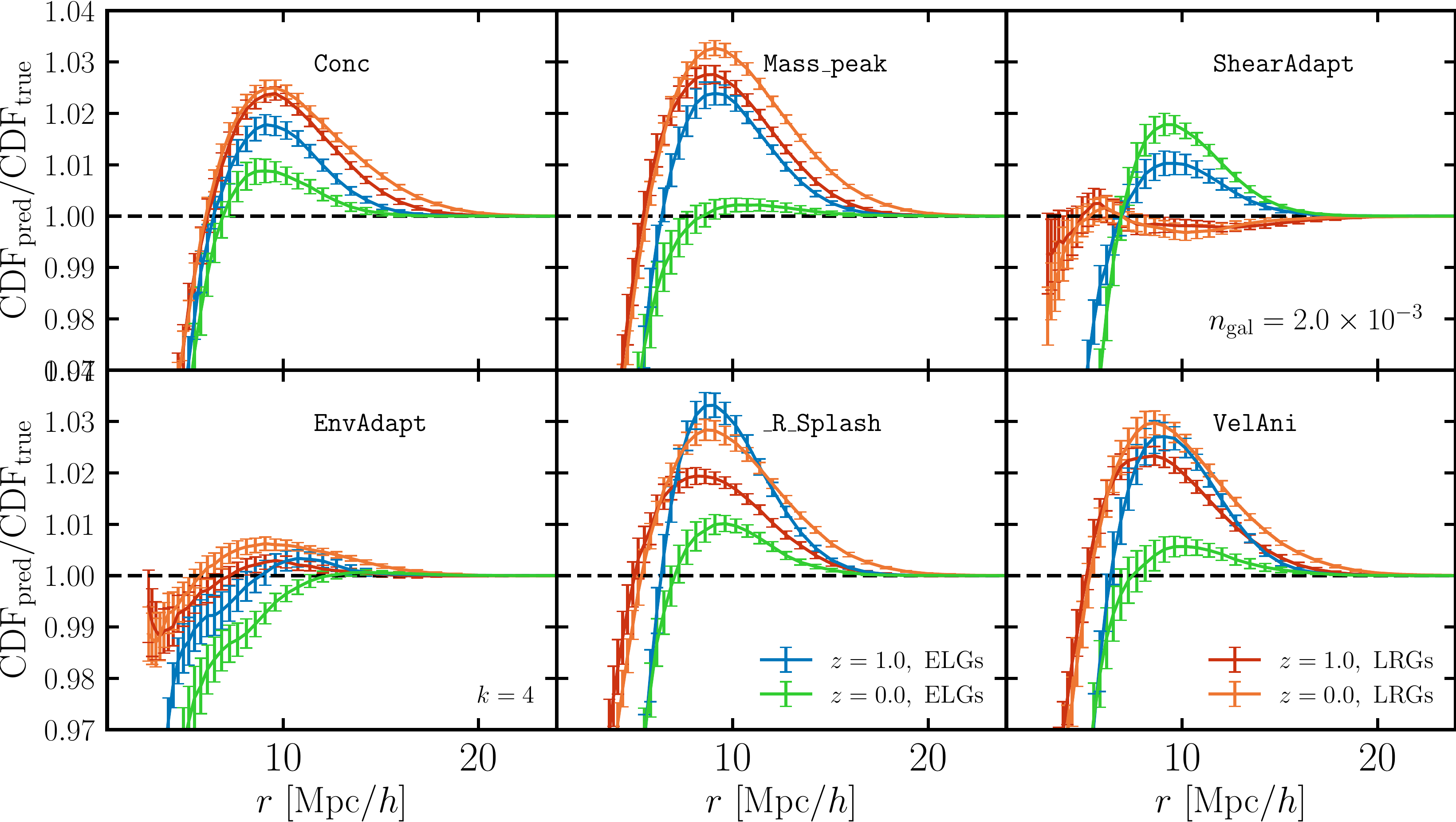}
\caption{Cumulative distribution function ratio between the predicted and ``true'' ``high''-density ELG and LRG samples at $z = 1$ and $z=0$. The order of the nearest-neighbour statistics shown here is $k = 4$. We also explore other orders ($k = 1$, $2$, and $8$), but find that our conclusions do not change qualitatively. The predicted samples are obtained via our extended halo occupation model (see Section~\ref{sec:halo_prop}) for secondary assembly bias, and each panel adopts a different halo property as its extension.  Interestingly, we find that the halo property that yields the best performance is ``environment'', for which the maximum value of the ratio is 1.01, which is lower than any of the other samples. Second best is the ``shear''-enhanced model. This shows that the nearest-neighbour counts indeed capture additional information about the galaxy distribution and as such can be used to study assembly bias and break degeneracies between halo occupation model parameters and cosmological parameters.}
\label{fig:kNN_ramp}
\end{figure*}

In Fig.~\ref{fig:kNN_ramp}, we compute the peaked kNN-CDF (CDF, see Section~\ref{sec:kNN} for definition) for our proposed halo occupation model augmented with secondary assembly bias (each panel adopts a different halo property as the extension). We note that when comparing the nearest neighbour statistics of two samples, it is necessary to match their number densities, so we randomly downsample either of our galaxy catalogues (predicted or ``true''), assigning equal weight to each galaxy. Typically the difference in the total galaxy number after downsampling is $\sim$1\%.

One can interpret deviations from unity in the plots as an excess or deficit of galaxies that are $k$-degree-separated from a random point in space at a specific scale $r$. We have also explored other orders, $k = 1$, $2$,  and $8$ (not shown), but find that our conclusions do not change qualitatively. Interestingly, we find that the halo property that yields the best performance is ``environment'', for which the maximum value of the ratio is lower than for any of the other samples. Second best is the ``shear''-enhanced model. This shows that the nearest-neighbour counts indeed capture additional information about the galaxy distribution and as such can be used to study assembly bias and break degeneracies between (and within) halo occupation model parameters and cosmological parameters.

Similarly to Fig.~\ref{fig:gab_ramp}, the intrinsic halo properties (concentration, peak mass, and velocity anisotropy) are more successful in reproducing the nearest-neighbour statistics of the ELG samples. We note that on smaller scales, $r \lesssim 5 \ \hMpc$, all models show substantial deviations, which is expected, as the value of the CDF function on these scales is dependent on the satellite model and close to zero, so the ratio blows up easily. Physically, contributions to these scales come from galaxies with very dense distributions that are typically found within a single halo. Thus, they are extremely sensitive to the three-dimensional distribution of galaxies within clusters.  The number of clusters we can probe is still quite paltry, however, as we are limited by the volume of the simulation. On the other end of the radial range, $r \gtrsim 20 \ \hMpc$, the CDF is sensitive to the galaxy distribution in void regions, which are also dependent on the simulation volume. It is, therefore, worth revisiting the question of how assembly bias affects these extreme scales with larger dark matter only simulations that have been equipped with a plausible mechanism for populating galaxies based on the predictions of high-fidelity hydro simulations such as MTNG740.

\section{Summary and Conclusions}
\label{sec:conc}

The goal of this study and our companion paper \citep{Hadzhiyska+2022a}  is to create a flexible augmented halo occupation prescription capable of recovering the clustering of MTNG galaxies down to very small scales, $r \sim 0.1 \ \hMpc$, and which can be readily adopted in the analysis of current and future surveys such as DESI, \textit{Euclid}, and \textit{Roman}. In this second paper, we have focused on the large-scale clustering and in particular, on determining the assembly bias signature of red and blue samples. We emphasize that we do not claim that the MTNG740 simulation that we use as our benchmark captures the complex physics of the real Universe perfectly, but rather that it serves as a good example of what assumptions might reasonably be questionable in galaxy-halo modeling.

We have introduced a novel method for incorporating assembly bias into the standard HOD framework (Section~\ref{sec:halo_prop}). Its main advantages are: (1) it adds only a small number of free parameters to the halo model (two per added halo property) and it is generalizable beyond tertiary assembly bias, (2) it models centrals and satellites separately, as they are governed by separate physical processes, (3) it is independent of the scale of variation of the halo properties used as assembly bias proxies, and (4) it is easily interpretable, as it appears as a correction to the mean halo occupation of centrals and satellites. In Section~\ref{sec:res}, we have tested the model in a robust and self-consistent manner by fitting it to the ``true'' occupations of haloes for different tracers and examining how well the occupation-fitted models predict a large-scale distribution of galaxies that agrees with the direct MTNG outputs.

Our main findings can be summarized as follows:
\begin{itemize}
    \item Both LRGs and ELGs exhibit substantial galaxy assembly bias ($r \gtrsim 1 \ \hMpc$) as shown in Fig.~\ref{fig:gab}. In particular, for all but the ELG population at $z = 0$, the mass-only HOD samples are less clustered than the ``true'' MTNG samples, indicating that the galaxies prefer to occupy more biased haloes than anticipated by the HOD model. The ELG sample at $z = 0$, however, has positive galaxy assembly bias, which signifies that the mass-only HOD puts galaxies in more biased haloes than the ``true'' ELG parents. In addition, the signal is scale-dependent, which has important implications for cosmological analysis, as it suggests that the linear bias approximation breaks down in the case of ELGs at fairly large scales ($r \sim 10 \ \hMpc$, compared with the typically assumed $r \sim 1 \ \hMpc$).
    \item In Fig.~\ref{fig:gab_mass}, we illustrate that the most significant contribution to the galaxy assembly bias signal comes from centrals in low-mass haloes, both for the LRGs and ELGs, but the satellite population has a more noticeable effect on the two-halo term for the ELGs rather than the LRGs. 
    \item Similarly to previous analyses \citep{1911.02610, 2021MNRAS.502.3599H, 2021MNRAS.502.3242X, 2021MNRAS.502.3582Y, 2021arXiv211102422D}, we show that the galaxy assembly bias gap can be closed by extrinsic halo properties (e.g., ``environment'') rather than intrinsic ones (e.g., concentration). However, these analyses have mostly been concerned with the red population of galaxies, while in this work we also add a blue population (ELGs) into the mix. In Fig.~\ref{fig:gab_ramp}, we show that the models enhanced with the halo property ``environment'' still perform well with the two red samples, but ``shear''-enhanced models are able to recover the clustering on large scales for all samples we consider. We conjecture that the galaxy formation process of the blue (ELG) sample may be dependent on the presence of massive clusters nearby, to which the property ``shear'' is more sensitive. This supports the idea of two-halo galaxy conformity on large scales explored also in Paper~I \citep{Hadzhiyska+2022a}. We also see that the clustering of the blue (ELG) populations is more sensitive to intrinsic halo properties than the red (LRG) populations.
    \item We explored the effect of redefining two important halo properties, mass and radius, on the clustering of galaxies, by adopting the physically intuitive ``splashback'' correction through an empirical relation proposed in \citet{2020ApJ...903...87D}. We showed in Fig.~\ref{fig:r_splash} that the splashback radius at $z = 1$ is on average nearly 1.5 times larger than the virial radius used in the rest of the paper. This finding has important implications for assembly bias. In the case of LRGs, it leads to a natural boost in the bias and in the one-halo to two-halo transition regime, which improves the agreement with the MTNG clustering data (see Fig.~\ref{fig:corr_splash}).
    \item In Fig.~\ref{fig:gab_plane}, we showed that there is slight improvement in the two-halo term when we augment our halo occupation model with tertiary assembly bias dependence. In particular, the model combining ``shear'' and ``environment'' does exhibit slightly better agreement with MTNG than each property separately. The finding that including tertiary assembly bias is not an essential requirement for reconciling the two-halo term of LRGs is welcome news for cosmologists who use HOD models to analyze galaxy survey data, since it saves the need to include more free parameters. For ELGs, it depends on the required accuracy and the available resources for the analysis whether a correction for tertiary assembly bias is worthwhile.
    \item Extrinsic halo properties (``environment,'' ``shear'') play a role in reducing the galaxy assembly bias signal, likely due to a combination of reasons, some of which are of astrophysical character and some of algorithmic. The algorithmic reasons involve issues such as
      percolation of halos with the FoF halo finder, which mostly affects the satellite occupation distribution, or how halo mass is defined, which affects both the mean central and satellite occupations. The astrophysical reasons are more difficult to pinpoint, but our tests provide some hints for an environmental dependence of star formation and for cooperative galaxy formation \citep[see Paper I][]{Hadzhiyska+2022a}, which might induce assembly bias in both the red and blue galaxy populations.
    \item Since real observations are done using the angular positions and redshifts of  galaxies, we have studied the galaxy assembly bias signal in redshift space through the quadrupole, $\xi_{\ell=2}$, which is sensitive to the peculiar velocity along the line of sight in addition to the true position of galaxies. Fig.~\ref{fig:xil2_ramp} illustrates the assembly bias effect in redshift space. We confirm that the ``shear''-augmented models perform best als in this case.   
    \item
 We have also explored a  higher-order statistics (Fig.~\ref{fig:kNN_ramp})
 based on nearest-neighbour counts (defined in Section~\ref{sec:kNN}) and find that the ``shear''-augmented model, which was best at matching the two-point clustering on large scales, exhibits a larger discrepancy than the ``environment''-based model. This demonstrates that higher-order statistics supply additional information about assembly bias and can be used to break degeneracies between the halo model parameters, and at the same time cautions that looking at two-point statistics alone can be misleading.
\end{itemize}

In this work and our companion paper \citep{Hadzhiyska+2022a},  we have suggested augmented HOD models that can reproduce the two-point galaxy clustering signals of a state-of-the-art hydrodynamical simulation of galaxy formation very well. We consider such models invaluable for the analysis of large galaxy surveys, and thus ultimately for attaining a profound understanding of the physics that governs our Universe. While our work has focused on making the HOD predictions for two-point galaxy statistics more accurate, and in particular to account for assembly bias, we have also seen that higher-order statistics may not necessarily be improved in lock-step, and in fact could prefer different secondary or tertiary parmeters. It will be an interesting question left for future study whether extensions of HOD models can perform as well for different higher-order statistics as they now do for the standard two-point statistics.

\section*{Acknowledgements}

The authors gratefully acknowledge the Gauss Centre for Supercomputing (GCS) for providing computing time on the GCS Supercomputer SuperMUC-NG at the Leibniz Supercomputing Centre (LRZ) in Garching, Germany, under project pn34mo. This work used the DiRAC@Durham facility managed by the Institute for Computational Cosmology on behalf of the STFC DiRAC HPC Facility, with equipment funded by BEIS capital funding via STFC capital grants ST/K00042X/1, ST/P002293/1, ST/R002371/1 and ST/S002502/1, Durham University and STFC operations grant ST/R000832/1. CH-A acknowledges support from the Excellence Cluster ORIGINS which is funded by the Deutsche Forschungsgemeinschaft (DFG, German Research Foundation) under Germany’s Excellence Strategy – EXC-2094 – 390783311. VS and LH acknowledge support by the Simons Collaboration on “Learning the Universe”. LH is supported by NSF grant AST-1815978.  SB is supported by the UK Research and Innovation (UKRI) Future Leaders Fellowship [grant number MR/V023381/1]. SC acknowledges the support of the ``Juan de la Cierva Incorporac\'ion'' fellowship (IJC2020-045705-I).

\section*{Data Availability}

The data underlying this article will be shared upon reasonable request to the corresponding 
authors. All MTNG simulations will be made publicly available in 2024.



\bibliographystyle{mnras}
\bibliography{example} 







\bsp	
\label{lastpage}
\end{document}